\documentclass[11pt,letterpaper]{article}
\usepackage[margin=1in]{geometry}

\usepackage{xspace}
\usepackage{xcolor}
\usepackage{amscd}
\usepackage{amsmath}
\usepackage{amssymb}
\usepackage{amstext}
\usepackage{amsthm}
\usepackage{thm-restate}
\usepackage{bm}
\usepackage{colonequals}
\usepackage{tikz}
\usepackage{booktabs}
\usepackage{array}
\usepackage{tabularx}
\usepackage{enumitem}
\usepackage[pagebackref=true,colorlinks,urlcolor=blue,linkcolor=purple,citecolor=blue,bookmarks,bookmarksopen,bookmarksnumbered]{hyperref}
\usepackage[capitalise,nameinlink]{cleveref}

\usepackage{mathtools}
\usepackage{amssymb, amsfonts, amsthm}

\usepackage{array}
\usepackage{verbatim}
\usepackage{mathrsfs}
\usepackage{graphicx}
\usepackage{algorithm}
\usepackage{algpseudocode}
\usepackage{caption}



\newcommand*{\alglineref}[1]{\hyperref[#1]{Line~\ref*{#1}}}

\usepackage{tikz}
\usetikzlibrary{decorations.pathreplacing}

\usepackage[toc,page]{appendix}

\usepackage{titlesec}
\titleformat{\subsubsection}[runin]
{\normalfont\normalsize\bfseries}
{\thesubsubsection}{1em}{}

\theoremstyle{plain}
\newtheorem{theorem}{Theorem}[section]
\newtheorem{lemma}[theorem]{Lemma}

\newtheorem{corollary}[theorem]{Corollary}

\theoremstyle{definition}
\newtheorem{definition}[theorem]{Definition}

\DeclareSymbolFont{bbold}{U}{bbold}{m}{n}
\DeclareSymbolFontAlphabet{\mathbbold}{bbold}

\newcommand{\poly}{\mathrm{poly}}

\renewcommand{\epsilon}{\varepsilon}

\newcommand{\AC}{\mathsf{AC}}
\newcommand{\SAC}{\mathsf{SAC}}
\newcommand{\Logspace}{\mathsf{L}}
\newcommand{\BPL}{\mathsf{BPL}}

\newcommand{\LocalRead}{\mathsf{LocalRead}}
\newcommand{\GapMaj}{\mathsf{GapMaj}}

\numberwithin{equation}{section}

\newif\ifshownotes
\shownotestrue

\ifshownotes
    \newcommand{\knote}[1]{{\color{cyan} (Kuan: #1)}}
	\newcommand{\hstnote}[1]{\textcolor{brown}{\textbf{[Shengtang: #1]}}}
    \newcommand{\wnote}[1]{\textcolor{purple}{\textbf{[Ruiyang: #1]}}}
\else
    \newcommand{\knote}[1]{}
	\newcommand{\hstnote}[1]{}
    \newcommand{\wnote}[1]{}
\fi

\title{High Precision Multiplicative-Error Approximation of $\#$SAC$^1$ Functions in Logspace Uniform AC$^1$}
\author{
Kuan Cheng\thanks{CFCS, School of Computer Science, Peking University. ckkcdh@pku.edu.cn.}
\and
Shengtang Huang\thanks{CFCS, School of Computer Science, Peking University. peanuttang1320061044@gmail.com.}
\and
Ruiyang Wu\thanks{CFCS, School of Computer Science, Peking University. wuruiyang@stu.pku.edu.cn.}
}
\date{}
\begin{document}

\maketitle
\begin{abstract}

We show that approximations of $\#\SAC^1$ functions  (i.e. functions computed by logspace-uniform polynomial-size positive semi-unbounded arithmetic circuits of depth $O(\log n)$) can be computed in logspace uniform $\AC^1$ (denoted as $\Logspace\text{-}\AC^1$) with multiplicative error $2^{-\poly\log n}$.
Previously, approximations of $\#\SAC^1$ functions were known
to be in logspace-uniform $\mathsf{TC}^1$ via exact computation
given by Allender, Jiao, Mahajan, and Vinay (TCS'98).

A main corollary of our result is that the following models beyond $\BPL$ are computable by  $\Logspace\text{-}\AC^1$ with multiplicative error $2^{-\poly\log n}$:
\begin{itemize}
    \item Probabilistic Auxiliary Push-down Machines (Probabilistic $\mathsf{AuxPDM}s$), which are polynomial time logspace Turing Machines (TMs) with an auxiliary stack having unbounded size and a read-once random tape;
    \item Read-$k$ Probabilistic TMs with polynomial time and logarithmic space, where $k$ is a constant and Read-$k$ means the random tape is two-way accessible and each cell can be visited for at most $k$ times. 
\end{itemize}
Notice that this is a direct generalization and improvement for $\BPL \subseteq \Logspace\text{-}\AC^1$ of Cheng and Wang (CCC' 24), since the classes of these computations are both super sets of the class of probabilistic logspace TMs, and the approximation error is improved from additive reverse polynomial to multiplicative reverse quasi-polynomial. 

$\mathsf{AuxPDM}$ was introduced by Cook (JACM'71) and its probabilistic version was first studied by Macarie and Ogihara (TCS'98). 
It is known that this model can compute strongly-few $s$-$t$ connectivity which is regarded as candidates to separate $\mathsf{BQL}$ from $\BPL$ by Apers and Edenhofer (CCC'25).
Read-$k$ Probabilistic logspace TMs was studied by Impagliazzo, Nisan, Wigderson (STOC'94) where a pseudorandom generator for this model was given. 

We mention that multiplicative error approximations are stronger than additive error approximations in the sense that they can even be used to decide non-deterministic computations.
Recently OpenAI announced that $\BPL = \Logspace$. 
Even if considering derandomizing probabilistic logspace, our $\Logspace\text{-}\AC^1$ computation has the advantage that the error can be $2^{-\poly\log n}$ and can be multiplicative.

\end{abstract}
\thispagestyle{empty}

\newpage
\setcounter{page}{1}

\section{Introduction}
\label{sec:intro}

Borodin, Cook, and Pippenger showed that an $S$-space probabilistic
Turing Machine (TM) admits an $O(S^2)$-time parallel simulation
\cite{BCP83}. At logarithmic space, this gives a classical
$O(\log^2 n)$ parallel-time benchmark, i.e. a logspace-uniform
$\mathsf{NC}^2$ simulation.
Cheng and Wang~\cite{ChengWang2024} proved that Probabilistic Logspace TMs can be simulated by logspace uniform $\AC^1$ (denoted as $\Logspace\text{-}\AC^1$) with reverse polynomial additive error, by developing a precision-boosting method for approximate matrix powering. 
It has been an interesting question whether computation models beyond probabilistic logspace TMs and/or higher precision simulations can also be done by $\Logspace\text{-}\AC^1$.
In this work, we consider the following more general computation models which are well-known in the literature.

Auxiliary Pushdown Machine ($\mathsf{AuxPDM}$) is introduced by Cook~\cite{Cook1971}. We denote $\mathsf{AuxPDM}$-$\mathsf{TISP}(T(n), S(n))$ as a Turing Machines (TM) with running time $T(n)$, work space $S(n)$ which has an auxiliary stack (unbounded size). We take $\mathsf{AuxPDM} =  \mathsf{AuxPDM}\text{-}\mathsf{TISP}(\poly(n), O(\log n) )$ by default.
In the literature $\mathsf{AuxPDM}$s are also denoted as Auxiliary Pushdown Automatas ($\mathsf{AuxPDA}$).
It is shown by Sudborough \cite{Sudborough1976AuxPDA} that $\mathsf{AuxPDM}$s are actually equivalent to LogDCFL which is the class of languages reducible to deterministic context-free languages under logspace many-to-one reductions.
Cook \cite{cook1979deterministic} also showed that $\mathsf{AuxPDM}$s can be computed by TMs with both polynomial time and $O(\log^2 n)$ space i.e. computable in SC$^2$.
Non-deterministic versions of $\mathsf{AuxPDM}$ and $\mathsf{AuxPDM}\text{-}\mathsf{TISP}(T(n), S(n))$ are denoted as $\mathsf{NAuxPDM}$ and $\mathsf{NAuxPDM}\text{-}\mathsf{TISP}(T(n), S(n))$ respectively.
Notice that we also take $\mathsf{NAuxPDM} = \mathsf{NAuxPDM}\text{-}\mathsf{TISP}(\poly(n), O(\log n) )$ by default.
The languages decidable by $\mathsf{NAuxPDM}$s are denoted as LogCFL which is the class of languages reducible to context-free languages under logspace many-to-one reductions.
We also consider its probabilistic version allowing two-sided errors.
Macarie and Ogihara studied probabilistic auxiliary pushdown automata in the unbounded-error setting
\cite{MacarieOgihara1998}. Venkateswaran subsequently considered its
bounded-error model \cite{Venkateswaran2009}.
\begin{definition}[Probabilistic  $\mathsf{AuxPDM}$s]
\label{def:intro-probabilistic-auxpdm}
   A probabilistic AuxPDM is a TM with an auxiliary unbounded stack, and an extra read-once random tape. Each step of the machine can push/pop one symbol to/from the top of the stack, or do nothing to the stack (we call this an idle operation on stack, or just idle for simplicity). 
\end{definition}
\begin{definition}[$\mathsf{BPAuxPDM}$]
\label{def:bpauxpdm}
The class $\mathsf{BPAuxPDM}\text{-}\mathsf{TISP}(T(n), S(n))$ consists of
all languages $L$ for which there is a probabilistic  auxiliary pushdown
machine $M$ that, on every input $x$ and every random tape,
runs in time $T(|x|)$ and uses $S(|x|)$ work-tape space.
The auxiliary stack is not counted toward the space bound.
For every input $x$,
\[
\begin{aligned}
x\in L &\implies \Pr_r[M(x;r)\text{ accepts}]\ge \frac23,\\
x\notin L &\implies \Pr_r[M(x;r)\text{ accepts}]\le \frac13.
\end{aligned}
\]
We denote $\mathsf{BPAuxPDM}$ for $\mathsf{BPAuxPDM}\text{-}\mathsf{TISP}(\poly(n),O(\log n))$.
\end{definition}
We note here that $\mathsf{AuxPDM}$ can compute certain problems that are candidates to separate more powerful classes from $\BPL$.
For example the strong-few s-t connectivity problem
STCON$_{sf} = \left\{ \langle G, s, t, 1^k \rangle \mid \forall i, j\in V(G), N(i, j)\le k, N(s,t)\ge 1   \right\}$ is considered by Apers and Edenhofer \cite{ApersEdenhofer2025} as a candidate separating Bounded-error Quantum logspace ($\mathsf{BQL}$) from $\BPL$, where $N(i, j)$ is a function counting the number of paths between $i$ and $j$. 
This problem is computable by $\mathsf{AuxPDM}$s \cite{ChengWu2026SC}.

Another interesting model is a direct generalization of $\BPL$, which allows polynomial-time logspace machines to access each random bit for at most $k$ times, where $k$ is a constant and the random tape is two-way accessible.
To be more accurate, the random-tape head moves only between adjacent cells and leaves each cell at most $k$ times during the computation.
We denote this class as $\LocalRead_k\text{-}\BPL$.
Notice that with two-way access, repeated reads of a cell must return the same bit, creating dependencies between visits~\cite{Nisan1993}. \
Impagliazzo, Nisan, and Wigderson~\cite{ImpagliazzoNisanWigderson1994} studied $\LocalRead_k\text{-}\BPL$ and gave a seminal pseudorandom generator for this class.
Cheng and Wu~\cite{ChengWu2026SC} obtained an $\mathsf{SC}^2$ simulation for this model.

In fact both models are closely related to a counting version of $\SAC^1$. $\SAC^i$ circuits are circuits consist of $\mathsf{AND}$, $\mathsf{OR}$ $\mathsf{NOT}$ gates, where $\mathsf{AND}$ gates only have fan-in two, the $\mathsf{OR}$ gates can have unbounded fan-in, and negations are allowed only at the inputs.
$\SAC^i$ circuits are closely related to $\mathsf{AuxPDM}$s, as shown by Venkateswaran \cite{Venkateswaran1991} that LogCFL is equivalent to logspace uniform $\SAC^1$. 
$\#\SAC^1$ is a counting version of $\SAC^1$. Originally it is defined via
$\#\SAC^1 := \#\mathsf{NAuxPDM}\text{-}\mathsf{TISP}(\poly(n),O(\log n))$ by Vinay \cite{Vinay1991}, which is the counting version of $\mathsf{NAuxPDM}$-$\mathsf{TISP}(\poly(n),O(\log n))$ i.e. asking for outputting the number of accepting computation paths.
Niedermeier and Rossmanith also investigated counting and unambiguous versions of auxiliary pushdown machine and semi-unbounded circuits \cite{NR95}. 
It is well-known that $\#\SAC^1$ can alternatively be defined via the following arithmetic model. 
\begin{definition}[Positive Semi-Unbounded Arithmetic Circuit \cite{Vinay1991, AllenderJiaoMahajanVinay1998}]
\label{def:SAC}
A positive semi-unbounded arithmetic circuit is a circuit over the semi-ring $(\mathbb{N},+,\cdot)$ (i.e. the semi-ring over natural numbers with additive operator ``$+$'' and multiplicative operator ``$\cdot$''), defined as the following.
The inputs are boolean variables $x_1, \ldots, x_n$ and can also have some fixed $0$s and $1$s.
The computation is over $\mathbb{N}$.
The circuit can flip the boolean variables immediately after the input, i.e. can compute $\bar{x}_j$ which is the Boolean complement of $x_j$.
For the remaining layers of computation, the circuits can only take additions (called addition gates) and multiplications (called multiplication gates) in $\mathbb{N}$.
The addition gates can have unbounded fan-in.
The multiplication gates can only have fan-in two.
All gates can have unbounded fan-out.
The final output of the circuit is an integer in $\mathbb{N}$.
\end{definition}
Arithmetic circuits are powerful, as one of the early striking result by Valiant, Skyum, Berkowitz, Rackoff \cite{ValiantSkyumBerkowitzRackoff1983} shows that, if a multivariate polynomial function has polynomial size and polynomial algebraic degree, over a semi-ring which is commutative, then there is another arithmetic circuit which can compute the same function over the same semi-ring while having polynomial size and depth  \(O(\log^2 n)\) where even both ``$+$'' and ``$\cdot$'' gates can have bounded fanin.
More generally,  ($\Logspace$-uniform) $\#\SAC^i$ is defined as (logspace-uniform) polynomial-size families of positive semi-unbounded arithmetic circuits of depth $O(\log^i n)$ (e.g. \cite{AllenderJiaoMahajanVinay1998}). 
Vinay showed that (logspace uniform) $\#\SAC^1$ can be computed in (logspace uniform) $\mathsf{NC}^2$~\cite{Vinay1991}. 
These functions can also be computed exactly by logspace-uniform
$\mathsf{TC}^1$ circuits~\cite[Theorem~4.2]{AllenderJiaoMahajanVinay1998}.
It remains a challenge whether such computation can be performed
by more restricted circuits, even approximately.

Another motivation of this work is about generalizing the techniques used in \cite{ChengWang2024}. The main technical tools of \cite{ChengWang2024} include an error reduction for matrix powering computable by $\Logspace\text{-}\AC^1$, and a low-depth multiplicative-error approximate counting process computable by low-depth $\Logspace\text{-}\AC$ circuits. 
The error reduction of \cite{ChengWang2024} is a variant of the error reduction for derandomizing probabilistic space bounded computations. 
We think this kind of error reduction can be viewed as having two main types. One is the recursive type initially given by Braverman, Cohen, Garg \cite{braverman2019pseudorandom}. The other is the Richardson iteration type initially given by Ahmadinejad, Kelner, Murtagh, Peebles, Sidford, and Vadhan \cite{ahmadinejadHighprecisionEstimationRandom2022}. A long sequence of work extensively studied and generalized this technique \cite{braverman2019pseudorandom, chattopadhyayOptimalErrorPseudodistributions2020, ahmadinejadHighprecisionEstimationRandom2022, cohenErrorReductionWeighted2021, hozaBetterPseudodistributionsDerandomization2021, pynePseudodistributionsThatBeat2021, cohenApproximatingIteratedMultiplication2023, puttermanOptimalDerandomizationMediumWidth2022, chenWeightedPseudorandomGenerators2023, chattopadhyayRecursiveErrorReduction2023, chengWuWeightedPseudorandomGenerators2026, chen25better, chenImprovedErrorReduction2026}.
The multiplicative-error approximate counting of \cite{ChengWang2024} is an extension of the approximate counting given by Ajtai \cite{ajtai1990approximate}. For both the error reduction technique and the low-depth multiplicative error counting technique, it is an interesting question whether they can be generalized to use in broader types of computations.

\subsection{Our Results}
\label{subsec:our-results}


Our main theorem shows that multiplicative-error approximations to $\#\SAC^1$ functions can be computed by $\Logspace$-uniform $\AC^1$.

\paragraph{Multiplicative-error approximation of $\#\SAC^1$ functions.}

\begin{theorem}[Multiplicative Approximation of Positive Arithmetic Circuits]
\label{thm:intro-positive-approximation}
Let $K(n),t(n)\ge1$ be logspace-computable, polynomially bounded integer-valued functions. Let $(F_n:\{0,1\}^n\to\mathbb{N}_{\ge0})_{n\ge1}$ be a function family computed by a polynomial-size, $\Logspace$-uniform family of positive semi-unbounded arithmetic circuits of depth at most $K(n)$. There is a polynomial-size, $\Logspace$-uniform Boolean $\AC$ circuit family of depth $O(K(n)+t(n))$ that, on every input $x\in\{0,1\}^n$, outputs a nonnegative number $\widetilde F_n(x)$ satisfying
\[
    (1-2^{-t(n)})F_n(x)
    \le \widetilde F_n(x)
    \le F_n(x).
\]
The output is represented with $t(n)+2$ significand bits and an $O(K(n)+\log n)$-bit exponent.
\footnote{All numerical approximations in our theorems are output in the floating-point representation (see \Cref{def:floating-point} for definition).}
\end{theorem}

For high-precision approximation, we further improve the depth dependence on the accuracy parameter $t(n)$.

\begin{theorem}[High-Precision Multiplicative Approximation]
\label{thm:intro-high-precision}
Let $K(n),t(n)\ge1$ be logspace-computable, polynomially bounded integer-valued functions. Let $(F_n:\{0,1\}^n\to\mathbb{N}_{\ge0})_{n\ge1}$ be a function family computed by a polynomial-size, $\Logspace$-uniform family of positive semi-unbounded arithmetic circuits of depth at most $K(n)$. There is a polynomial-size, $\Logspace$-uniform Boolean $\AC$ circuit family of depth
\[
O\left( K(n) + \frac{\log n}{\log \log n} \log \left( 2 + \frac{t(n)}{K(n)} \right) \right)
\]
that, on every input $x\in\{0,1\}^n$, outputs a nonnegative number $\widetilde F_n(x)$ satisfying
\[
    (1-2^{-t(n)})F_n(x)
    \le \widetilde F_n(x)
    \le F_n(x).
\]
The output is represented with $t(n)+3$ significand bits and an $O(K(n)+\log n)$-bit exponent.
\end{theorem}
We remark that the simulation can also be done if the description of the target circuit is given as an input. See e.g. our \cref{cor:input-circuit-approximation}.
We also remark that one may try to get a better asymptotic bound for depth dependence on $K$ and $t$. But we are not intended to do that as our main purpose is to show that the approximation of $\#\SAC^1$ is computable by $\Logspace$-uniform $\AC^1$ circuits with multiplicative reverse quasi-polynomial error. 

Taking $K(n)=\Theta(\log n)$ in
\Cref{thm:intro-high-precision} gives the following corollary.

\begin{corollary}
\label{cor:intro-sac1-approximation}
For every logspace-computable integer-valued function $t(n)$ with $1\le t(n) \le \poly\log n$, every function family in $\Logspace$-$\#\SAC^1$ admits a multiplicative $(1-2^{-t(n)})$-approximation from below computable by a $\Logspace$-uniform $\AC^1$ circuit family.
\end{corollary}

\paragraph{Simulating probabilistic auxiliary pushdown machines.}


By combining Vinay's counting characterization~\cite{Vinay1991}
with our high-precision approximation theorem, we obtain the
following result.

\begin{theorem}[Approximating Acceptance Probabilities of Auxiliary Pushdown Machines]
\label{thm:intro-auxpdm-approximation}
Let $M$ be a polynomial-time probabilistic auxiliary pushdown machine with $O(\log n)$ work space and a read-once random tape, and let $p_M(x)$ denote its acceptance probability. For every logspace-computable integer-valued function $1 \le t(n) \le \poly \log n$, there is a polynomial-size, $\Logspace$-uniform $\AC^1$ circuit family that, on every input $x\in\{0,1\}^n$, outputs a nonnegative number $\widetilde p_M(x)$ satisfying
\[
    (1-2^{-t(n)})p_M(x)
    \le \widetilde p_M(x)
    \le p_M(x).
\]
The output is represented with $t(n)+3$ significand bits and an $O(\log n)$-bit exponent.
\end{theorem}

\begin{corollary}
\label{cor:intro-auxpdm-containment}
\[
    \mathsf{BPAuxPDM}\text{-}\mathsf{TISP}(\poly(n),O(\log n))
    \subseteq \Logspace\text{-}\AC^1.
\]
\end{corollary}
Note that our theorem can be viewed as a direct improvement to $\BPL \subseteq \Logspace\text{-}\AC^1$, since a probabilistic $\mathsf{AuxPDM}$ has a polynomial size stack in addition to its $O(\log n)$ work space,  and the approximation error is improved from reverse polynomial to reverse quasi-polynomial.

\paragraph{Simulating logspace machines with bounded random-tape visits.}

By developing a reduction from $\LocalRead_k\text{-}\BPL$ to approximation of $\#\SAC^1$ functions, we attain the following result.

\begin{theorem}[Approximating Acceptance Probabilities with Bounded Random-tape Visits]
\label{thm:intro-local-read-approximation}
Fix an integer $k\ge1$. Let $M$ be a polynomial-time, logarithmic-space locally read-$k$ machine, and let $p_M(x)$ denote its acceptance probability. For every logspace-computable integer-valued function $1 \le t(n) \le \poly \log n $, there is a polynomial-size, $\Logspace$-uniform $\AC^1$ circuit family that, on every input $x\in\{0,1\}^n$, outputs a nonnegative number $\widetilde p_M(x)$ satisfying
\[
    (1-2^{-t(n)})p_M(x)
    \le \widetilde p_M(x)
    \le p_M(x).
\]
The output is represented with $t(n)+3$ significand bits and an $O(\log n)$-bit exponent. Moreover, the acceptance-count function
\[
    F_M(x)=
    \#\{r\in\{0,1\}^{R(n)}:M(x;r)\text{ accepts}\}
\]
belongs to $\#\Logspace$ (see \Cref{def:sharp-logspace}), where $R(n)$ is the length of the random tape.
\end{theorem}

\begin{corollary}
\label{cor:intro-local-read-containment}
For every fixed integer $k\ge1$,
\[
    \LocalRead_k\text{-}\BPL
    \subseteq \Logspace\text{-}\AC^1.
\]
\end{corollary}
Again our theorem can be viewed as a direct improvement to $\BPL \subseteq \Logspace\text{-}\AC^1$, since our read-k model allows multiple two-way access to the random tape, and the approximation error is improved from reverse polynomial to reverse quasi-polynomial.
We note that if one directly apply the PRG of Impagliazzo, Nisan, and Wigderson~\cite{ImpagliazzoNisanWigderson1994} then as their PRG has seed length $O(\log^2 n)$ when $k$ is a constant, it only yields quasipolynomial-size circuits.
We note that Cheng and Wu~\cite{ChengWu2026SC} obtained an $\mathsf{SC}^2$ simulation for this model. 

We also note that there are other types of read-$k$ models considered in the literature which are of independent interests.
David, Nguyen, Papakonstantinou,  Sidiropoulos \cite{DNPS11}  also studied probabilistic $\mathsf{AuxPDM}$s whose
access to the random tape is limited to a specified number
of complete passes \cite{DNPS11}. This gives another way
to restrict randomness in the presence of a stack. 
Bogdanov, Papakonstantinou, and Wan further constructed a
pseudorandom generator for such models
\cite{BPW12}. 
Gurjar and Volk constructed a pseudorandom generator
for read-$k$ oblivious branching programs whose reading
sequence is known in advance \cite{GV20}.
Doron and Lang \cite{DL26} recently attained a new construction improving \cite{GV20}.


\paragraph{About  $\mathsf{L}=\mathsf{BPL}$} We note that an OpenAI preprint has announced  $\mathsf{L}=\mathsf{BPL}$ \cite{OpenAI26Logspace}. We emphasize that our theorems have the advantage that they give reverse quasi-polynomial multiplicative error estimates of acceptance probabilities for the two more general models mentioned above.

\subsection{Technical Overview}
\label{subsec:technical-overview}
We show our main result by giving a new precision-boosting (error reduction) framework specifically for relative approximation of positive arithmetic circuits. 
Notice that we are considering a more general model than the substochastic matrix powering considered by \cite{ChengWang2024} and our new framework will be significantly different from theirs. 
But the difficulty here is somewhat similar to that of \cite{ChengWang2024}, i.e. the target arithmetic circuit already has $O(\log n)$ depth. If we want to compute a single layer with high precision, then that already may cost roughly $O(\log n)$ layers of $\AC$ circuits. The precision-boosting framework of \cite{ChengWang2024} overcome this barrier by somehow using low precision approximations for long matrix powering together with high precision approximations for short matrix powering to attain high precision computations for long matrix powering with small overall depth of $\AC$ circuit computation. It is an interesting question whether this technology as well as other types of error reduction can work for more general target models.


We will describe our main techniques for \cref{thm:intro-positive-approximation} \cref{thm:intro-high-precision} \cref{thm:intro-auxpdm-approximation} \cref{thm:intro-local-read-approximation}. Throughout, vector and matrix inequalities are interpreted entrywise, e.g., $A\le B$ means $A_{ij}\le B_{ij}$ for every $i,j$.
\paragraph{Correcting errors across layers.}
We prove \cref{thm:intro-positive-approximation} for logspace-computable, polynomially bounded arithmetic depth $K$ and accuracy parameter $t_0$.

Our goal is to obtain a more accurate approximation at a target layer using approximate values from several earlier layers.


Let $x_i$ be the vector of gate outputs at layer $i$ on a fixed input. For $1\le i\le K$, let $\Phi_i$ be the map computed by layer $i$ from the outputs of layer $i-1$, so that $x_i=\Phi_i(x_{i-1})$. Each coordinate of $\Phi_i$ is a polynomial of degree at most two with nonnegative coefficients. Fix $1\le\ell\le k\le K$ and set $a=k-\ell$. Suppose that we have recursively computed nonnegative lower approximations $y_i$ for $a\le i<k$. At each interior layer, we can compare the approximation $y_i$ with the value $\Phi_i(y_{i-1})$ computed from the preceding approximation. Their difference, $\delta_i=\Phi_i(y_{i-1})-y_i$, relates the unknown errors at these two layers: writing $e_i=x_i-y_i$ and expanding $\Phi_i$ around $y_{i-1}$ (i.e. Taylor Expansion) gives
\[
    e_i=\delta_i+J_i e_{i-1}+Q_i(e_{i-1}).
\]
Here $D$ denotes the Jacobian operator, $J_i=D\Phi_i(y_{i-1})$ gives the linear term $J_i e_{i-1}$, and $Q_i(e_{i-1})$ collects the terms quadratic in the input error. Both $J_i$ and $\delta_i$ are computable from the approximations. We use the linear part of this recurrence to carry each $\delta_i$ to layer $k$, and add these contributions to the direct approximation $\Phi_k(y_{k-1})$ for $x_k$:
\[
    C=\Phi_k(y_{k-1})+
    \sum_{i=a+1}^{k-1}J_k\cdots J_{i+1}\delta_i.
\]

Expanding the recurrence shows that the correction cancels the first-order errors from the interior layers. The error $x_k-C$ therefore consists only of the propagated boundary error $e_a$ and the propagated quadratic remainders. These terms are nonnegative, so $C\le x_k$. If the interval starts at the exact input layer, then $e_a=0$ and only the quadratic remainders remain.

We can bound the relative error of $C$ in terms of the relative errors of the recursive approximations $y_i$. If $e_{i-1}\le\eta x_{i-1}$, positivity gives $Q_i(e_{i-1})\le\eta^2x_i$, so the quadratic remainder has relative size at most $\eta^2$. Moreover, positivity and the degree bound give $J_i x_{i-1}\le 2x_i$, limiting the amplification during propagation to a factor of two per layer. These bounds determine the recursive precisions needed to control both the boundary error and the sum of the propagated quadratic remainders.

\paragraph{Obtaining Boolean depth $O(K+t_0)$.}
We construct $b_k(t)$, a lower approximation to $x_k$ with relative error at most $2^{-t}$, using recursive subcircuits followed by one correction circuit. The recursive subcircuits supply $y_{k-j}=b_{k-j}(p_j)$, where
\[
    \ell=\min\{k,\lceil t/2\rceil+3\},
    \qquad
    p_j=\lceil t/2\rceil+2j
    \quad(1\le j\le\ell).
\]
The choice of $\ell$ ensures that the interval either reaches the input layer or extends far enough to keep the boundary error within the error budget.

To evaluate the correction from the recursive approximations, we substitute $\delta_i=\Phi_i(y_{i-1})-y_i$ into the formula for $C$. We collect all added terms into $C^+$ and all subtracted terms into $C^-$, so that $C=C^+-C^-$. Since the Jacobians and the approximate values are nonnegative, both expressions involve only sums and products of nonnegative quantities.
Once the recursive approximations are available, we form the Jacobians in parallel and compute their products by balanced matrix multiplication. This gives positive arithmetic circuits for $C^+$ and $C^-$ of depth $h=O(\log(\ell+1))$. Since each expression is at most $2^\ell x_k$, relative accuracy $2^{-q}$ with $q=t+\ell+O(1)$ suffices to keep the error in their difference within the remaining budget. Our shallow evaluator combines the approximate-counting circuits of Cheng and Wang~\cite[Theorem~14]{ChengWang2024} with exact arithmetic on polynomial-length words and achieves this accuracy in Boolean depth
\[
    O\left(h\left(1+\frac{\min\{q+h,\log n\}}{\log\log n}\right)\right).
\]
Here $q=O(t+1)$ and $h=O(\log(t+2))$, so this depth is $O(t+1)$ for every polynomially bounded recursive accuracy parameter $t$. We compute a lower approximation to $C^+$ and an upper approximation to $C^-$, then clip their difference at zero and round downward to obtain $b_k(t)$.

To bound the depth of the full construction, we account for the corrections along each recursive dependency path. The recursive subcircuits run in parallel, so the depth of $b_k(t)$ is the maximum depth of its recursive inputs plus the local correction cost $O(t+1)$. Each recursive step from $(k,t)$ to $(k-j,p_j)$ decreases the potential $4k+t$ by $2j+\lfloor t/2\rfloor=\Omega(t+1)$. Thus the local cost is at most a constant times this decrease. Along any dependency path starting at $(K,t_0)$, the decreases sum to at most $4K+t_0$, giving total Boolean depth $O(K+t_0)$.

To control size, we reuse the subcircuit for each approximation wherever it is needed. Every recursive pair $(k,s)$ satisfies $4k+s\le 4K+t_0$, so there are at most $O((K+1)(K+t_0))$ distinct pairs, which is polynomial in $n$. Their significands have $O(K+t_0)$ bits. If $N$ bounds the gates and wires of the layered arithmetic circuit, the exact outputs through layer $i$ are bounded by $V_i$, where $V_0=2$ and $V_i\le NV_{i-1}^2$. Hence $\log V_K=O(2^K\log n)$. Every nonzero recursive approximation lies in $[1/2,V_K]$, so its exponent requires $O(K+\log n)$ bits. Thus the local evaluators operate on polynomial-length encodings. Each local correction circuit has polynomial size, and sharing one subcircuit per pair therefore gives polynomial total size. The parameters and wiring can all be generated in logarithmic space. The final output $b_K(t_0)$ has $t_0+2$ significand bits and an $O(K+\log n)$-bit exponent.


\paragraph{Refining to higher precision.}
For logspace-computable, polynomially bounded integers $K,t\ge1$, (in \Cref{sec:error-reduction}) we reduces the relative error to $2^{-t}$. 
Let $X$ denote the vector of gate outputs across all layers on the fixed input. 
W.l.o.g. we assume that for each gate, all its inputs are from the immediate previous layer.
The computation at each gate expresses its output as a polynomial in the outputs of its input gates; collecting these polynomials defines a map $\Phi$ satisfying $X=\Phi(X)$. 
Since addition gates are linear and multiplication gates have fan-in two, the function $\Phi$ has the equation $\Phi(z)=c+Lz+Q(z), \forall z$, where $Q$ is homogeneous quadratic.
We first apply \Cref{thm:intro-positive-approximation} to obtain an initial approximation $y$ of $X$, with relative error $2^{-CK}$, for a sufficiently large constant $C$, in Boolean depth $O(K)$. 
Then we implement a small adjustment to let the initial approximation $y$ satisfy $0\le y\le\Phi(y)\le X$.

The residual $r=\Phi(y)-y$ is computable from $y$, whereas
the error $e=X-y$ depends on the unknown vector $X$.
Write $J(z)=D\Phi(z)$ and abbreviate $J=J(y)$.
Since $X=\Phi(X)$ and $Q$ is homogeneous quadratic,
\[
e=\Phi(y+e)-y=r+Je+Q(e).
\]
Ignoring the quadratic term suggests solving $(I-J)e=r$
to estimate the error.
Since each layer depends only on the preceding layer,
$J^{K+1}=0$.
Thus the inverse
\[
G=(I-J)^{-1}=\sum_{i=0}^KJ^i
\]
exists and is nonnegative.
The resulting estimate of the error is $Gr$.
Adding $Gr$ to $y$ leaves precisely the quadratic remainder:
\[
X-(y+Gr)
=G((I-J)e-r)
=GQ(e).
\]
We use $y+Gr$ as a better approximation for $X$.
If $0\le X-y\le\alpha X$, then
$Q(X-y)\le\alpha^2Q(X)\le\alpha^2X$.
The above identity and the bound $GX\le HX$, where
$H=2^{K+1}-1$, therefore give
$0\le X-(y+Gr)\le H\alpha^2X$. 
Because $H$ is exponential in $K$, constant initial accuracy does not ensure contraction; this is why we need the $\Theta(K)$ accurate bits in the initial approximation.

The ideal correction requires an inverse that changes with $y$.
Recomputing this inverse in every round would not give
the desired depth bound.
We therefore maintain an approximation $R$ to $G$ alongside $y$.

After correcting $y$ to $y'=y+Rr$, we need to approximate
the new inverse $G'=(I-J(y'))^{-1}$.
As with the correction to $y$, we estimate the unknown error
$G'-R$ from a computable residual.
Define $B=I-(I-J(y'))R$.
The identity $G'-R=G'B$ shows that the exact correction
would be $G'B$.
Since $G'$ is unknown, we replace it by $R$ in the correction:
\[
R'=R+RB,
\qquad
G'-R'=(G'-R)B=G'B^2.
\]
Equivalently, the new inverse residual is
$I-(I-J(y'))R'=B^2$.
The update therefore squares the inverse residual using
only matrix sums and products.

Changing $y$ also changes the target inverse.
Our joint analysis controls both the gate-value error
and the inverse residual, accounting for the change
in the Jacobian.
Starting from sufficiently accurate approximations,
the updates attain relative gate-value error $2^{-t}$
within $O(\log(2+t/K))$ rounds.

Using \Cref{lem:shallow-positive-circuit}, we approximate the
damped inverse \((I-(1-\eta)J(y))^{-1}\) from below with relative
factor \(1-\eta\), where \(\eta=2^{-6K-20}\). By
\Cref{lem:er-init}, the resulting matrix \(R\) has a nonnegative
residual \(I-(I-J(y))R\) that is sufficiently small. The
initialization can be computed by Boolean circuits of depth
\(O(K+\log n/\log\log n)\). Each refinement round uses constantly many exact matrix operations on polynomial-length dyadic rationals. We can show that this costs $O(\log n/\log\log n)$ $\AC$-circuit depth to compute. 
So overall, we obtain relative error $2^{-t}$ with polynomial size logspace-uniform $\AC$ circuits of depth
\[
O\left(K+\frac{\log n}{\log\log n}\log\left(2+\frac{t}{K}\right)\right).
\]
For $K=O(\log n)$, every fixed polylogarithmic precision is therefore attainable in $\Logspace$-uniform $\AC^1$.


To approximate the acceptance probabilities of both machine
models, we express the number of accepting random strings of each
model as a $\Logspace$-uniform $\#\SAC^1$ function. We then
approximate this number with high precision and divide by the
corresponding total number of possible random strings.

\paragraph{Counting computations with a pushdown tape.}
Let \(M\) be a polynomial-time probabilistic auxiliary pushdown
machine with logarithmic work space and a read-once random tape.
By padding with ignored random bits, we may assume that every
computation uses exactly \(T(n)\) random bits, where \(T(n)\) is a
logspace-computable polynomial bound.
Since the random tape is read once, each random string corresponds
to exactly one nondeterministic path; replacing each random-bit
choice by a nondeterministic binary choice therefore yields a
machine whose accepting-path count is
\[
 F_M(x)
 =
 \#\{r\in\{0,1\}^{T(n)}:M(x;r)\text{ accepts}\}.
\]
By Vinay's characterization~\cite{Vinay1991}
,
\(F_M\) is computed by a \(\Logspace\)-uniform \(\#\SAC^1\)
circuit family.
Theorem~\ref{thm:intro-high-precision} then gives a
\(\Logspace\)-uniform \(\AC^1\) circuit family that outputs
\(\widetilde F_M(x)\) satisfying
\[
 (1-2^{-t(n)})F_M(x)\le \widetilde F_M(x)\le F_M(x)
\]
for every logspace-computable integer-valued function \(t\) with
\(1\le t(n)\le\operatorname{polylog}n\).
Since \(p_M(x)=2^{-T(n)}F_M(x)\), rescaling by \(2^{-T(n)}\)
preserves the relative error and yields the corresponding
approximation to \(p_M(x)\).
Constant $t$ therefore suffices for the bounded-error
simulation.

\paragraph{Counting computations with bounded random-tape visits.}
We first construct a nondeterministic logspace verifier whose number of accepting paths equals the count of accepting random strings of a machine with a locally read-$k$ random tape, then convert this count into a $\Logspace$-uniform $\#\SAC^1$ circuit so that our approximation theorem applies.

The verifier represents a computation by recording its visits to each random-tape cell. An \emph{episode} is the computation during one visit to a cell, ending with the next random-head move or a halt. The cell's bit and the starting configuration determine the entire episode. We number the episodes $1,2,\ldots$ in chronological order and use a \emph{record} for each cell consisting of its bit and a list of triples $(s,c,\alpha)$, where $s$ is the number of an episode at that cell, $c$ is its starting configuration and $\alpha$ records arrival from the left or right,
or marks the initial episode. The entries are listed in increasing order of $s$. The locally read-$k$ condition gives at most $k+1$ entries per record. Since each entry uses $O(\log n)$ bits, the entire record uses $O(\log n)$ bits for fixed $k$.

To check that the records describe an accepting computation, the verifier guesses them from left to right and checks how the episodes follow one another. For each entry $(s,c,\alpha)$, it simulates the episode from configuration $c$ using the cell's recorded bit. If the random head moves to a neighboring cell in configuration $c'$, the verifier requires that cell's record to contain $(s+1,c',\alpha')$, where $\alpha'$ records arrival from the current cell, matching the two entries as consecutive episodes. It also requires a unique episode numbered $1$ starting in the prescribed initial configuration, exactly one matching predecessor for every other episode, and a matching successor for every episode except the unique one that halts in acceptance. The increasing episode numbers rule out cycles, while the predecessor and successor conditions ensure that all entries belong to a single complete accepting computation. The arrival label indicates whether the verifier should look for
the preceding episode in the left neighboring cell's record or in
the right one's. The verifier can therefore perform these checks while retaining only two adjacent records at a time, using logarithmic space.

Each random tape on which the original machine accepts uniquely determines the computation and its ordered records. Thus the verifier has exactly one accepting path for each such tape, placing the acceptance-count function in $\#\Logspace$. Counting paths in its configuration graph by balanced matrix multiplication gives the required $\Logspace$-uniform $\#\SAC^1$ circuit. Our approximation theorem then gives a relative approximation to the count, and hence the same relative accuracy for the acceptance probability.
\section{Preliminaries}
\label{sec:preli}

\paragraph{Some notations.} Throughout the paper, all logarithms are base two. Inequalities between vectors are interpreted componentwise. We write $\mathbb N=\{0,1,2,\ldots\}$.
For all the computation models considered in this work, their input lengths are denoted by $n$, and constants hidden in asymptotic big-O notations may depend on the computation models, but not on $n$.
An integer-valued function $f(n)$ is logspace-computable if its binary representation can be computed from $1^n$ in deterministic $O(\log n)$ space, and polynomially bounded if $|f(n)|\le n^{O(1)}$.  Vector and matrix inequalities are interpreted entrywise:
$u\le v$ means $u_i\le v_i$ for every $i$, and
$A\le B$ means $A_{ij}\le B_{ij}$ for every $i,j$.

\subsection{Boolean $\AC$ Circuits and Uniformity}
\label{subsec:boolean-uniformity}

We begin with the Boolean circuit model and the uniformity convention used throughout the paper.

\begin{definition}[Boolean $\AC$ Circuit]
A Boolean $\AC$ circuit consists of input gates, Boolean constants, $\mathsf{NOT}$ gates, unbounded-fan-in $\mathsf{AND}$ and $\mathsf{OR}$ gates, and one or more designated output gates. The \emph{size} of a circuit is the number of gates and wires, and its \emph{depth} is the maximum number of gates on a path from an input gate to an output gate.
\end{definition}

\begin{definition}[$\Logspace$-Uniform Circuit Family]
A circuit family $(C_n)_{n\ge1}$ is \emph{$\Logspace$-uniform} if there is a deterministic Turing machine that, on input $1^n$, outputs a description of $C_n$ using $O(\log n)$ work space.
\end{definition}

A $\Logspace$-uniform $\AC^1$ circuit family is a $\Logspace$-uniform family of polynomial-size Boolean $\AC$ circuits of depth $O(\log n)$. We write $\Logspace$-$\AC^1$ for the class of Boolean functions, or equivalently languages, computed by $\Logspace$-uniform $\AC^1$ circuit families. We also use Boolean circuits with multiple output gates to represent numerical values.

\subsection{Positive Arithmetic Circuits}
\label{subsec:positive-arithmetic-circuits}

We next introduce the positive arithmetic circuit model used in our approximation theorems and applications.

\begin{definition}[Positive Semi-Unbounded Arithmetic Circuit \cite{Vinay1991, AllenderJiaoMahajanVinay1998}]
A positive semi-unbounded arithmetic circuit is a circuit over the semi-ring $(\mathbb{N},+,\cdot)$ defined as the following.
The inputs are boolean variables $x_1, \ldots, x_n$ and can also have some fixed $0$s and $1$s.
The computation is over $\mathbb{N}$.
The circuit can flip the boolean variables immediately after the input, i.e. can compute $\bar{x}_j$ which is the Boolean complement of $x_j$.
For the remaining layers of computation, the circuits can only use addition gate ``$+$'' and multiplication gate ``$\cdot$'' in $\mathbb{N}$.
The addition gates can have unbounded fan-in.
The multiplication gates can only have fan-in two. All gates can have unbounded fan-out.
\end{definition}


A $\Logspace$-uniform $\#\SAC^1$ circuit family is a $\Logspace$-uniform, polynomial-size family of positive semi-unbounded arithmetic circuits of depth $O(\log n)$ \cite{AllenderJiaoMahajanVinay1998}. We write $\Logspace$-$\#\SAC^1$ for the class of nonnegative integer-valued functions computed by $\Logspace$-uniform $\#\SAC^1$ circuit families.

\subsection{Floating-Point Representation and Arithmetic}
\label{subsec:floating-point-arithmetic}

We use a binary floating-point representation for nonnegative dyadic rationals.

\begin{definition}[Floating-Point Representation]
\label{def:floating-point}
A nonnegative real number $x$ is called a $(p,b)$-number if it admits a representation consisting of a significand $M$, an exponent $e$, and a zero flag. If the zero flag is set, then $x=0$. Otherwise,
\[
x=M2^{e-\ell},\qquad 2^{\ell-1}\le M<2^\ell,\qquad 2\nmid M,
\]
where $\ell\le p$ is the binary length of $M$, and $e\in\mathbb Z$ has a signed binary representation of at most $b$ bits, including the sign bit.

The representation of $x$ consists of the binary representations of $M$, $e$, and the zero flag. We call $p$ the \emph{precision bound} and $b$ the \emph{exponent-length bound}.
\end{definition}

We use addition and multiplication in their usual senses and write $[A-B]_+ := \max\{A-B,0\}$ for subtraction of nonnegative numbers. For $s\ge1$, let $\operatorname{RoundDown}_s(x)$ and $\operatorname{RoundUp}_s(x)$ be the largest and smallest nonnegative numbers, respectively, with at most $s$ significand bits that are no greater and no smaller than $x$. They satisfy
\[
0\le x-\operatorname{RoundDown}_s(x)\le2^{1-s}x,\qquad 0\le\operatorname{RoundUp}_s(x)-x\le2^{1-s}x.
\]

We record four numerical lemmas used in our circuit constructions. The proofs of them are deferred to \Cref{sec:numerical-primitives}. In the rest of this subsection, $m, p, b, q \ge 1$ are all logspace-computable, polynomially bounded integer-valued functions of $n$.

We first begin with elementary operations.

\begin{restatable}{lemma}{elementaryArithmetic}
\label{lem:elementary-arithmetic}
Given nonnegative $(p,b)$-numbers $A,B$, the following operations admit polynomial-size, $\Logspace$-uniform Boolean $\AC$ circuits of constant depth:
\begin{enumerate}
\item $\operatorname{RoundDown}_q(A)$ and $\operatorname{RoundUp}_q(A)$.
\item Exact addition $A+B$ and subtraction $[A-B]_+$ when $b=O(\log n)$.
\item Exact multiplication $AB$ when $p=O(\log n)$.
\end{enumerate}
\end{restatable}

The exponent restriction in exact addition and subtraction cannot be dropped. For example, $2^{2^n}$ and $1$ have short floating-point representations, but their exact difference $2^{2^n}-1$ requires exponentially many significand bits. We can avoid this difficulty by rounding the result.

\begin{restatable}{lemma}{roundedAddSub}
\label{lem:rounded-add-sub}
Given two nonnegative $(p,b)$-numbers $A,B$, polynomial-size, $\Logspace$-uniform Boolean $\AC$ circuits of constant depth compute
\[
\operatorname{RoundDown}_q(A+B),\quad \operatorname{RoundUp}_q(A+B),\quad
\operatorname{RoundDown}_q([A-B]_+),\quad \operatorname{RoundUp}_q([A-B]_+).
\]
Each output is an $(q,b+O(\log n))$-number.
\end{restatable}

For high-precision computation, we also need exact products with long significands and exact sums of many terms.

\begin{restatable}{lemma}{exactArithmetic}
\label{lem:er-arithmetic}
The following computations admit polynomial-size, $\Logspace$-uniform Boolean $\AC$ circuits of depth $O(\log n/\log\log n)$:
\begin{enumerate}
\item Given two nonnegative $(p,b)$-numbers $A,B$, compute their exact product $AB$ as a $(2p,b+2)$-number.
\item If $b=O(\log n)$, given $m$ nonnegative $(p,b)$-numbers $A_1,\ldots,A_m$, compute their exact sum $S=\sum_{i=1}^m A_i$ as a $(p',b')$-number, where $p'=p+2^b+\lceil\log m\rceil+2$ and $b'=b+O(\log\log n)$.
\end{enumerate}
\end{restatable}

Finally, we consider approximate summation without restrictions on the exponent length.

\begin{restatable}{lemma}{positiveSummation}
\label{lem:positive-sum}
There is a polynomial-size, $\Logspace$-uniform Boolean $\AC$ circuit family of depth
\[
O\left(1+\frac{\min\{q,\log n\}}{\log\log n}\right)
\]
that takes as input nonnegative $(p,b)$-numbers $A_1,\ldots,A_m$ and outputs two nonnegative $(q+6,b+O(\log\log n))$-numbers $L,U$ satisfying
\[
(1-2^{-q})S\le L\le S\le U\le(1+2^{-q})S,\qquad S=\sum_{i=1}^m A_i.
\]
\end{restatable}
\section{Multiplicative Approximation of Positive Arithmetic Circuits}


We restate \Cref{thm:intro-positive-approximation}
and prove it in this section.

\begin{theorem}[Multiplicative Approximation of Positive Arithmetic Circuits]
\label{thm:approx-sharp-sac}
$K(n),t(n)\ge1$ are logspace-computable, polynomially bounded integer-valued functions. Let $(F_n:\{0,1\}^n\to\mathbb{N}_{\ge0})_{n\ge1}$ be a function family computed by a polynomial-size, $\Logspace$-uniform family $(G_n)_{n\ge1}$ of positive semi-unbounded arithmetic circuits of depth at most $K(n)$.

Then there is a polynomial-size, $\Logspace$-uniform Boolean $\AC$ circuit family of depth $O(K(n)+t(n))$ that, on every input $x\in\{0,1\}^n$, outputs a $(t(n)+2, O(K(n)+\log n))$-number $\widetilde F_n(x)$ in floating-point representation, satisfying
\[
    (1-2^{-t(n)})F_n(x)
    \le \widetilde F_n(x)
    \le F_n(x).
\]
In particular, if $K(n),t(n)=O(\log n)$, this approximation is computable by a $\Logspace$-uniform $\AC^1$ circuit family.
\end{theorem}

In \Cref{subsec:low-depth-evaluation}, we use the numerical primitives established earlier to approximate the outputs of shallow positive arithmetic circuits. We then use this evaluator in \Cref{subsec:approximation-error} to construct recursive approximations of the gate values, and in \Cref{subsec:approximation-complexity} we prove the depth, size, and uniformity bounds.

\subsection{Low-Depth Evaluation of Positive Arithmetic Circuits}
\label{subsec:low-depth-evaluation}

We first extend the numerical primitives in \Cref{subsec:floating-point-arithmetic} to the evaluation of shallow positive arithmetic circuits. In this subsection, input gates and arithmetic constants may carry arbitrary nonnegative $(p,b)$-numbers, rather than only Boolean values.
The following lemma can do multiplicative-error approximations for positive semi-unbounded arithmetic circuits of depth $h$, via $\Logspace\text{-}\AC$ circuits but with depths significantly larger than our target. We will use it later as a basic component in our final construction.  

\begin{lemma}
\label{lem:shallow-positive-circuit}
\label{lem:positive-expression-evaluator}
\label{lem:polynomial-encoding-evaluation}
Let $p, b, q, h \ge 1$  be logspace-computable, polynomially bounded integer-valued functions of $n$. Let $(H_n)_{n\ge4}$ be a polynomial-size, $\Logspace$-uniform family of positive semi-unbounded arithmetic circuits of depth at most $h=h(n)$. Suppose that its inputs and arithmetic constants are nonnegative $(p,b)$-numbers. Then there is a polynomial-size, $\Logspace$-uniform Boolean $\AC$ circuit family of depth
\[
    O\left(h\left(1+\frac{\min\{q+h,\log n\}}{\log\log n}\right)\right)
\]
that, for every output value $z$ of $H_n$, produces nonnegative $(q+h+9,O(b+h+\log\log n))$-numbers $L,U$ such that
\[
    (1-2^{-q})z\le L\le z\le U\le(1+2^{-q})z.
\]
\end{lemma}

\begin{proof}
Fix $n$ and write $H=H_n$. Set $r=q+h+3$ and $\theta=2^{-r}$. For every gate $v$ of $H$, with exact value $z_v$, we compute nonnegative lower and upper approximations $L_v,U_v$ with significand lengths at most $r+6$.

\par\smallskip\noindent\textit{(1) Description of circuits.} For input or constant gate $v$, let $L_v=\operatorname{RoundDown}_{r+6}(z_v)$ and $U_v=\operatorname{RoundUp}_{r+6}(z_v)$. For addition gate $v$ with children $v_1,\ldots,v_m$, apply \Cref{lem:positive-sum} in parallel to $\sum_i L_{v_i}$ and $\sum_i U_{v_i}$, taking their lower and upper outputs, respectively, with accuracy parameter $r$. For multiplication gate $v$ with children $v_1,v_2$, put
\[
    L_v=\operatorname{RoundDown}_{r+6}(L_{v_1}L_{v_2}),\qquad
    U_v=\operatorname{RoundUp}_{r+6}(U_{v_1}U_{v_2}).
\]

\par\smallskip\noindent\textit{(2) Analysis of errors.} All these approximations have at most $r+6$ significand bits. By \Cref{lem:positive-sum}, each addition introduces at most a factor $1\pm\theta$ of relative error, while the directed-rounding bound $2^{1-(r+6)}\le\theta$ gives the same guarantee at input and multiplication gates.

We claim that for every gate $v$ of depth at most $d$,
\begin{equation}
    (1-\theta)^{2^{d+1}-1}z_v\le L_v\le z_v\le U_v\le(1+\theta)^{2^{d+1}-1}z_v.
    \label{eq:shallow-evaluator-induction}
\end{equation}
We show that this follows from directed rounding. Suppose it holds for the children of a gate of depth $d\ge1$. At an addition gate, positivity and \Cref{lem:positive-sum} give
\[
\begin{aligned}
    L_v&\ge(1-\theta)\sum_i L_{v_i}\ge(1-\theta)^{2^d}z_v,\\
    U_v&\le(1+\theta)\sum_i U_{v_i}\le(1+\theta)^{2^d}z_v.
\end{aligned}
\]
For every multiplication gate, the product of the two child bounds followed by directed rounding gives the same inequalities with exponent $2(2^d-1)+1=2^{d+1}-1$. This proves \eqref{eq:shallow-evaluator-induction} by induction.

Now we use \eqref{eq:shallow-evaluator-induction} to show the final approximation bounds. Let $m_0=2^{h+1}-1$. Since $m_0\theta<2^{-q-2}\le1/8$, Bernoulli's inequality and $e^u\le1+2u$ for $0\le u\le1/8$ imply
\[
    (1-\theta)^{m_0}\ge1-m_0\theta\ge1-2^{-q},\qquad
    (1+\theta)^{m_0}\le e^{m_0\theta}\le1+2m_0\theta\le1+2^{-q}.
\]
Thus the required approximation bounds follows. Also notice that since $q\ge1$, a positive $z_v$ has $L_v\ge z_v/2>0$, and zero-valued gates have $L_v=U_v=0$; hence zero is detected exactly.

\par\smallskip\noindent\textit{(3) Exponent lengths.} We next bound the exponent lengths. Let $N\ge2$ be a polynomial bound on the number of gates and wires of $H$. A nonzero input $(p,b)$-number lies in $[2^{-A},2^A]$ for $A=2^b+1$. Each sum is at most $N$ times its largest summand. Each multiplication has its exponent length to be at most the addition of the logarithms of the factors. Induction on arithmetic depth therefore places all nonzero gate values in $[2^{-B},2^B]$, where $B=2^h(A+\log N)$. The approximations lie within a factor of two of the corresponding positive gate values, so their exponents have $O(b+h+\log\log n)$ bits. We use a common logspace-computable bound of this length at all gates.

\par\smallskip\noindent\textit{(4) Boolean depth.} Finally, we analyze Boolean depth. If $r\le\lceil\log n\rceil$, the multiplication and directed rounding at each gate have constant depth by \Cref{lem:elementary-arithmetic}, while \Cref{lem:positive-sum} implements addition in depth $O(1+r/\log\log n)$. If $r>\lceil\log n\rceil$, exact multiplication has depth $O(\log n/\log\log n)$ by \Cref{lem:er-arithmetic}, directed rounding has constant depth by \Cref{lem:elementary-arithmetic}, and addition has depth $O(\log n/\log\log n)$ by \Cref{lem:positive-sum}. Hence each arithmetic layer is implemented in Boolean depth
\[
    O\left(1+\frac{\min\{r,\log n\}}{\log\log n}\right).
\]
Since $r=q+h+3$, evaluating all $h$ layers takes the claimed depth. Every gate replacement has polynomial size, and the circuit is $\Logspace$-uniform by the logspace-computable parameters and the uniformity of \Cref{lem:elementary-arithmetic}, \Cref{lem:er-arithmetic} and \Cref{lem:positive-sum}. All outputs have at most $r+6=q+h+9$ significand bits and the exponent bound proved above.
\end{proof}

\subsection{Recursive Approximation and Error Analysis}
\label{subsec:approximation-error}

To express circuit evaluation as a sequence of polynomial maps,
we first put \(G_n\) into an equivalent layered form.
Let \(K=K(n)\), and let \(m\) be the number of gates of \(G_n\).
For every gate \(v\) of \(G_n\), we introduce a gate \(v^{(i)}\)
in each layer \(i=0,\ldots,K\).

If \(v\) is an input gate, its copies are defined as follows.
If \(v\) computes a literal \(x_j\) or \(\bar{x}_j\), then
\(v^{(0)}\) is an input gate for the same literal.
If \(v\) is a constant gate with value \(c\in\{0,1\}\), then
\(v^{(0)}\) is a constant gate with value \(c\).
For each \(1\le i\le K\), \(v^{(i)}\) is a fan-in-one
addition gate whose sole input is \(v^{(i-1)}\).

If \(v\) is an addition or multiplication gate with input list
\(u_1,\ldots,u_d\), then \(v^{(0)}\) is a constant-zero gate.
For each \(1\le i\le K\), \(v^{(i)}\) has the same gate type
as \(v\) and input list \(u_1^{(i-1)},\ldots,u_d^{(i-1)}\).
The output gate is \(o^{(K)}\), where \(o\) is the output
gate of \(G_n\).

Each layer has width \(m\), and each gate in layer \(i\ge1\)
receives inputs only from layer \(i-1\).
By induction on \(i\), \(v^{(i)}\) computes the same function
as \(v\) whenever \(i\) is at least the depth of \(v\).
Thus the new circuit computes the same function as \(G_n\).
The construction preserves \(\Logspace\)-uniformity and
increases the size by at most a factor of \(O(K)\). We continue to denote the resulting circuit by \(G_n\).

For a fixed Boolean input \(x\), let
\[
    x_0,x_1,\ldots,x_K\in\mathbb{N}^m
\]
be the vectors of gate values in layers \(0,\ldots,K\),
with coordinates indexed by the gates of \(G_n\).
Here \(x_0=x_0(x)\).
The gates in each layer \(i\ge1\) define a polynomial map
\[
    \Phi_i:\mathbb{R}_{\ge0}^m\longrightarrow\mathbb{R}_{\ge0}^m,
    \qquad
    x_i=\Phi_i(x_{i-1}),
    \qquad 1\le i\le K.
\]
Each coordinate of \(\Phi_i\) is a sum of coordinates
(possibly a single coordinate) or a product of two coordinates.
Hence each coordinate polynomial has degree at most two
and nonnegative coefficients.

For $0\le k\le K$ and every integer $t\ge1$, we will construct an approximation vector $b_k(t)$ satisfying
\[
    (1-2^{-t})x_k\le b_k(t)\le x_k.
\]
We will define $b_k(t)$ by induction on $k$, simultaneously for all positive integers $t$, starting from $b_0(t)=x_0$.

\paragraph{Construction.}
\label{par:parameter-selection}
Consider a target layer $k \in [1,  K]$ and an accuracy parameter $t\ge1$. Set
\[
    u=\lceil t/2\rceil,
    \qquad
    \ell=\min\{k,u+3\},
    \qquad
    p_j=u+2j\quad(1\le j\le\ell).
\]

Put $a=k-\ell$ and let
\[
    y_{k-j}=b_{k-j}(p_j),
    \qquad 1\le j\le\ell.
\]
The vectors $y_a,\ldots,y_{k-1}$ are available by induction. When $a=0$, we have $y_0=x_0$ exactly.

For $a+1\le i\le k$, define
\[
    q_i=\Phi_i(y_{i-1}),
    \qquad
    J_i=D\Phi_i(y_{i-1}),
\]
Here $J_i$ is the Jacobian matrix of $\Phi_i$ at $y_{i-1}$. For $a+1\le i<k$, define the defect
\[
    \delta_i =q_i-y_i.
\]

One can think that the defect $\delta_i$ measures the discrepancy between exactly computing the $i$-th layer using $y_{i-1}$, and our approximate computing of the $i$-th layer using $y_{i-1}$.

Now we define a main approximation quantity $C$ as the following, which can be thought of as propagating the defects in layers $a+1 , a+2 , \ldots, k-1 $ through subsequent Jacobians to attain a correction to $q_k$.
\begin{equation}
    C=q_k+
    \sum_{i=a+1}^{k-1}
    J_kJ_{k-1}\cdots J_{i+1}\delta_i.
    \label{eq:multilayer-correction}
\end{equation}

We can write $C=C^+-C^-$ as the following such that $C^+$ and $C^-$ can both be computed by positive semi-unbounded arithmetic circuits,
\begin{align}
    C^+
    &=q_k+
      \sum_{i=a+1}^{k-1}
      J_kJ_{k-1}\cdots J_{i+1}q_i,
      \label{eq:positive-correction-part}\\
    C^-
    &=\sum_{i=a+1}^{k-1}
      J_kJ_{k-1}\cdots J_{i+1}y_i.
      \label{eq:negative-correction-part}
\end{align}
Both $C^+$ and $C^-$ are positive arithmetic expressions in the available approximation vectors. Set the evaluation accuracy parameter to
\[
    q=t+\ell+3.
\]
Use \cref{lem:shallow-positive-circuit} to compute a lower approximation $L^+$ of $C^+$ and an upper approximation $U^-$ of $C^-$ satisfying
\[
    (1-2^{-q})C^+\le L^+\le C^+,
    \qquad
    C^-\le U^-\le(1+2^{-q})C^-.
\]
Set the output word precision to
\[
    s=t+2.
\]
i.e. only keep the first $s$ significant bits of the output.
We define $b_k(t)$ by
\begin{equation}
    b_k(t)
    =\operatorname{RoundDown}_s\bigl([L^+-U^-]_+\bigr).
    \label{eq:rounded-corrected-value}
\end{equation}

\paragraph{Correctness.}

We calculate the accuracy of the construction.

\begin{lemma}
\label{lem:multilayer-boosting}
For every $0\le k\le K$ and every integer $t\ge1$, the approximation defined above satisfies
\[
    (1-2^{-t})x_k\le b_k(t)\le x_k,
\]
\end{lemma}

\begin{proof}
We prove by induction on $k$. The case $k=0$ holds because $b_0(t)=x_0$ for every $t\ge 1$. Fix $k\ge1$ and $t\ge1$, and assume the claim holds at every earlier layer for all positive integer accuracy parameters. Then
\[
    (1-2^{-p_j})x_{k-j}\le y_{k-j}\le x_{k-j},
    \qquad 1\le j\le\ell.
\]

Write $e_i=x_i-y_i$ for $a\le i<k$, and, for $a+1\le i\le k$. We can canonically decompose $\Phi_i(z)$ to be
\[
    \Phi_i(z)=c_i+L_i(z)+Q_i(z),
\]
where $L_i$ and $Q_i$ are the homogeneous linear and quadratic parts
in $z$, respectively. All coefficients are nonnegative since the
underlying arithmetic circuit is positive. 
Using $x_{i-1}=y_{i-1}+e_{i-1}$, the Taylor expansion of $\Phi_i$ at $y_{i-1}$ gives
\begin{equation}\label{eq:extraxi}
\begin{aligned}
     x_i=\Phi_i(x_{i-1})
    &=\Phi_i(y_{i-1}+e_{i-1})\\
    &= \Phi_i(y_{i-1}) + D\Phi_i(y_{i-1}) e_{i-1} + \frac12 D^2\Phi_i(y_{i-1})(e_{i-1}\otimes e_{i-1})\\
    &=q_i+J_ie_{i-1}+Q_i(e_{i-1}).
\end{aligned}
\end{equation}
For $a+1\le i<k$, subtracting $y_i$ from both sides of the above equality yields
\begin{equation}\label{eq:extraei}
    e_i=\delta_i+J_ie_{i-1}+Q_i(e_{i-1}).
\end{equation}
Now consider the expansion for $x_k$. 
We start by using \eqref{eq:extraxi} to get
\[
    x_k = q_k+J_ke_{k-1}+Q_k(e_{k-1}).
\]
Then we substitute the $e_{k-1}$ of the Jacobian linear term $J_ke_{k-1}$ by the expansion of $e_{k-1}$ from \eqref{eq:extraei}.
Then we keep doing this, i.e. iteratively substituting the $e_{i}$s of the Jacobian linear terms by the corresponding expansion from \eqref{eq:extraei}, until the Jacobian linear term has $e_a$ inside. Notice that we don't substitute these $e_i$s in the quadratic terms.

In this way, we obtain
\[
\begin{aligned}
x_k
={}&q_k+\sum_{i=a+1}^{k-1}
J_k\cdots J_{i+1}\delta_i
+J_k\cdots J_{a+1}e_a\\
&+\sum_{i=a+1}^{k}
J_k\cdots J_{i+1}Q_i(e_{i-1}),
\end{aligned}
\]
By \eqref{eq:multilayer-correction}, the first two terms equal $C$.
Hence,
\begin{equation}
    x_k-C
    =J_k\cdots J_{a+1}e_a
     +\sum_{i=a+1}^{k}
       J_k\cdots J_{i+1}Q_i(e_{i-1}).
    \label{eq:correction-error-identity}
\end{equation}
Every term on the right is nonnegative. 

To bound these terms, we first control $J_k\cdots J_{i+1}$ for $a \le i \le k $. Notice that the entries of $D\Phi_i$ are polynomials with nonnegative coefficients.
Also recall that $y_{i-1}\le x_{i-1}$.
Hence $J_i=D\Phi_i(y_{i-1})\le D\Phi_i(x_{i-1})$. 
Therefore as $x_i$s are non-negative, we have
\[
\begin{aligned}
    J_ix_{i-1}
    &\le D\Phi_i(x_{i-1})x_{i-1}\\
    &=L_i(x_{i-1})+2Q_i(x_{i-1})\\
    &\le2\Phi_i(x_{i-1})=2x_i.
\end{aligned}
\]
Iteratively using this bound, we have for $a\le i\le k$,
\begin{equation}
    J_kJ_{k-1}\cdots J_{i+1}x_i
    \le2^{k-i}x_k.
    \label{eq:jacobian-product-bound}
\end{equation}

Recall that $a=k-\ell$, $\ell=\min\{k,u+3\}$.
If $a=0$, then by our construction we have $e_a=0$. So for this case the first term in \eqref{eq:correction-error-identity} vanishes. Otherwise, we have $\ell=u+3$. 
Now using \eqref{eq:jacobian-product-bound} taking $i = a$ and using the induction hypothesis, we have
\begin{equation}
\label{eq:ea}
    0\le J_k\cdots J_{a+1}e_a
    \le2^{\ell-p_\ell}x_k
    =2^{-2u-3}x_k
    \le2^{-t-3}x_k,
\end{equation}
where we also use $0\leq e_a= x_{a}-y_a\leq 2^{-p_\ell} x_a$, $p_\ell=u+2\ell$, and $2u\ge t$.

Now we bound $J_k\cdots J_{i+1}Q_i(e_{i-1})$. We denote $i=k-j+1$. The inductive hypothesis gives $0\le e_{i-1}\le2^{-p_j}x_{i-1}$. By the nonnegative coefficients and quadratic homogeneity of $Q_i$, as well as the definition of $\Phi_i$, we have
\begin{equation}
        0\le Q_i(e_{i-1})
    \le2^{-2p_j}Q_i(x_{i-1})
    \le2^{-2p_j}\Phi_i(x_{i-1})
    =2^{-2p_j}x_i.
\end{equation}

Now taking Jacobians on $Q_i(e_{i-1})$ and applying \eqref{eq:jacobian-product-bound}, we have
\begin{equation}
\label{eq:eqi1}
\begin{aligned}
    0
    &\le J_k\cdots J_{k-j+2}Q_{k-j+1}(e_{k-j})\\
    &\le2^{-2p_j}J_k\cdots J_{k-j+2}x_{k-j+1}\\
    &\le2^{j-1-2p_j}x_k.
\end{aligned}
\end{equation}

Since $p_j=u+2j$ and $2u\ge t$,
\begin{equation}
    \label{eq:eqi2}
    \sum_{j=1}^{\ell}2^{j-1-2p_j}
    =2^{-2u-1}\sum_{j=1}^{\ell}2^{-3j}
    \le\frac{2^{-t-1}}7.
\end{equation}
Combining \eqref{eq:ea}, \eqref{eq:eqi1} and \eqref{eq:eqi2} with \eqref{eq:correction-error-identity}, we obtain
\[
    0\le x_k-C
    \le\left(2^{-t-3}+\frac{2^{-t-1}}7\right)x_k.
\]

We next bound the additional error from numerical evaluation and rounding. Since $q_i=\Phi_i(y_{i-1})\le x_i$ and $y_i\le x_i$, applying \eqref{eq:jacobian-product-bound} to \eqref{eq:positive-correction-part} and \eqref{eq:negative-correction-part} gives
\[
    C^+\le(2^\ell-1)x_k,
    \qquad
    C^-\le(2^\ell-2)x_k.
\]
The guarantees for $L^+$ and $U^-$, together with $q=t+\ell+3$, therefore imply
\begin{equation}
\label{eq:eLU}
    \begin{aligned}
    0\le C-(L^+-U^-)
    &\le2^{-q}(C^++C^-)\\
    &\le(2^{\ell+1}-3)2^{-q}x_k\\
    &\le2^{-t-2}x_k.
\end{aligned}
\end{equation}

Put $d=[L^+-U^-]_+$. Since $L^+-U^-\le C\le x_k$ and $x_k\ge0$, we have $0\le d\le x_k$. Moreover, $d\ge L^+-U^-$, so
\[
    0\le x_k-d
    \le(x_k-C)+\bigl(C-(L^+-U^-)\bigr).
\]
Finally, as $b_k(t)=\operatorname{RoundDown}_s(d)$ with $s=t+2$, we have
\[
    0\le d-b_k(t)\le2^{1-s}d\le2^{-t-1}x_k.
\]
Adding all these error bounds together yields
\[
    0\le x_k-b_k(t)
    \le\left(
        2^{-t-3}+\frac{2^{-t-1}}7
        +2^{-t-2}+2^{-t-1}
       \right)x_k
    \le2^{-t}x_k.
\]

This completes the induction.
\end{proof}

\subsection{Depth, Size, and Uniformity}
\label{subsec:approximation-complexity}

We now analyze the Boolean circuits implementing the recursive construction in \Cref{subsec:approximation-error}. Let $K=K(n)\le n^a$ be the logspace-computable depth bound used in the layered construction, and fix a logspace-computable output accuracy parameter $1\le t_0=t_0(n)\le n^b$, where $b>0$ is fixed. Write $A(k,t)$ for the Boolean circuit computing $b_k(t)$ from $x_0$. The desired approximation is computed by $A(K,t_0)$.

\paragraph{Parameter and encoding length bounds.}
To bound the parameters occurring in the recursion, define
\[
    \Psi(k,t)=4k+t,
    \qquad
    P=\Psi(K,t_0)=4K+t_0.
\]
The computation of \(b_k(t)\) uses \(b_{k-j}(p_j)\)
for \(1\le j\le\ell\).
For each such \(j\), the choice
\(p_j=\lceil t/2\rceil+2j\) gives
\begin{equation}
    \Psi(k,t)-\Psi(k-j,p_j)
    =2j+\lfloor t/2\rfloor
    \ge\frac{t+1}{2}.
    \label{eq:approximation-potential-drop}
\end{equation}
Thus every pair $(k,t)$ reached from $(K,t_0)$ satisfies $4k+t\le P$, where $P=O(K+t_0)$ is polynomially bounded. Since $\ell\le\lceil t/2\rceil+3$, the parameters $p_j$, $q=t+\ell+3$, and $s=t+2$ are $O(t+1)$ and hence $O(P)$. In particular, the recursive inputs $b_{k-j}(p_j)$ have polynomially bounded precision $p_j+2$.

We next bound the exponents of the recursive inputs. Let $N\ge2$ be a polynomial bound on the number of gates and wires in the layered arithmetic circuit, and put
\[
    V_i=\max\{2,\|x_0\|_\infty,\ldots,\|x_i\|_\infty\}.
\]
The input bits and arithmetic constants are $0$ or $1$, so $V_0=2$. Each arithmetic layer satisfies $V_i\le N V_{i-1}^2$, and hence
\[
    \log V_K\le 2^K+(2^K-1)\log N.
\]
Since the exact gate values are nonnegative integers bounded by $V_K$, \Cref{lem:multilayer-boosting} implies that every nonzero coordinate of a recursive input lies in $[1/2,V_K]$. Its exponent has absolute value at most $1+\log V_K$ and can be represented using $O(K+\log\log n)$ bits. Hence every recursive input $b_{k-j}(p_j)$ is a $(p_j+2,O(K+\log n))$-number, as required by \Cref{lem:shallow-positive-circuit}.

We use a common logspace-computable exponent width of
$O(K+\log n)$ bits for all recursive outputs.

\paragraph{Depth.}
Fix a pair $(k,t)$ with $k\ge1$, and regard $y_a,\ldots,y_{k-1}$ as inputs. We construct positive semi-unbounded arithmetic circuits for $C^+$ and $C^-$ in \eqref{eq:positive-correction-part} and \eqref{eq:negative-correction-part} as follows. First, we compute all $q_i$ and $J_i$ in constant arithmetic depth. We then compute the matrix products $J_k\cdots J_{i+1}$ in parallel, using a balanced tree of matrix multiplications for each product, before applying them to the corresponding vectors and summing the results.

Each matrix multiplication or matrix--vector multiplication has constant arithmetic depth. Each matrix product involves at most $\ell$ factors, so the corresponding balanced tree has depth $O(\log(\ell+1))$. The final matrix--vector multiplications and summation add only constant depth, giving total arithmetic depth
\[
    h=O(\log(\ell+1)).
\]

Since $\ell=O(t+1)$, we have $h=O(\log(t+2))$ and $q+h=O(t+1)$. By \Cref{lem:shallow-positive-circuit}, the directed approximations $L^+$ and $U^-$ can be computed in parallel with relative error at most $2^{-q}$ and Boolean depth
\[
    O\left(\log(t+2)\left(1+\frac{\min\{t+1,\log n\}}{\log\log n}\right)\right)
    =O(t+1).
\]
For $t\le\log n$, this follows from $\log(t+2)=O(\log\log n)$ and $\log(t+2)=O(t+1)$. For $t>\log n$, use $\log(t+2)/(t+1)=O(\log\log n/\log n)$. The numerical inputs to this local computation are $(O(t+1),O(K+\log n))$-numbers. By \Cref{lem:rounded-add-sub}, the subtraction followed by directed rounding in \eqref{eq:rounded-corrected-value} has constant Boolean depth.

Let $D(k,t)$ denote the depth of $A(k,t)$. The recursive inputs are computed in parallel, followed by the local computation just described. Thus, for a constant $c$,
\begin{equation}
    D(k,t)\le
    \max_{1\le j\le\ell}D(k-j,p_j)+c(t+1),
    \qquad D(0,t)=O(1).
    \label{eq:approximation-depth-recurrence}
\end{equation}
For each \(1\le j\le\ell\),
\eqref{eq:approximation-potential-drop} gives
\[
    c(t+1)\le 2c\bigl(\Psi(k,t)-\Psi(k-j,p_j)\bigr).
\]
Along any recursive dependency path starting at $(K,t_0)$, the potential differences sum to at most $P$. The depth recurrence \eqref{eq:approximation-depth-recurrence} therefore gives
\[
    D(K,t_0)\le2cP+O(1)=O(K+t_0).
\]

\paragraph{Size and uniformity.}
To obtain polynomial size, we construct one local circuit for each pair in
\[
    \mathcal I=\{(k,t):0\le k\le K,\ 1\le t\le P-4k\},
\]
and share its output wherever the same approximation is needed. Each recursive input is connected to the corresponding output for $(k-j,p_j)$, while the base cases use the exact encodings of $b_0(t)=x_0$. Since $|\mathcal I|\le(K+1)P=O((K+1)(K+t_0))=n^{O(1)}$ and all local circuits satisfy a common polynomial size bound, the total size is polynomial.

For uniformity, index each Boolean gate by its pair $(k,t)$ and its index within the local circuit. A logspace transducer enumerates $\mathcal I$, computes the local parameters, and generates each local circuit from the uniform description of the layered arithmetic circuit using \Cref{lem:shallow-positive-circuit} and \Cref{lem:rounded-add-sub}. The balanced matrix products can be indexed by factor intervals and matrix coordinates. Although the numerical words may have polynomial length, their bit positions and all gate indices have $O(\log n)$ bits. The connections between local circuits are computable in logspace. Thus the resulting family is $\Logspace$-uniform.

\begin{proof}[Proof of \Cref{thm:approx-sharp-sac}]
Set $t_0=t(n)$, and let $\widetilde F_n(x)$ be the designated coordinate of $b_K(t_0)$. By \Cref{lem:multilayer-boosting}, for every input $x$,
\[
    (1-2^{-t_0})F_n(x)\le\widetilde F_n(x)\le F_n(x).
\]
The preceding analysis gives a polynomial-size, $\Logspace$-uniform Boolean $\AC$ circuit family computing this approximation with depth $O(K+t_0)$. Its output is a $(t_0+2,O(K+\log n))$-number, as required. When $K,t_0=O(\log n)$, these bounds give the stated $\Logspace$-uniform $\AC^1$ approximation.
\end{proof}

\section{AC$^1$ Simulation of Probabilistic Auxiliary Pushdown Machines}
\label{sec:stack}

We apply our approximation theorem to the acceptance probabilities
of probabilistic auxiliary pushdown machines, using the model of
\Cref{def:intro-probabilistic-auxpdm}
and~\Cref{def:bpauxpdm}. Only the work tapes count toward the space
bound, and the stated time and work-space bounds hold for every
random tape.

We first recall the counting version of nondeterministic auxiliary
pushdown machines and Vinay's arithmetic-circuit characterization.

\begin{definition}[Counting Auxiliary Pushdown Machines]
\label{def:counting-auxpdm}
For a nondeterministic auxiliary pushdown machine $N$, let
$\#\mathrm{acc}_N(x)$ denote the number of accepting computation
paths of $N$ on input $x$. The class
\[
 \#\mathsf{NAuxPDM}\text{-}\mathsf{TISP}
 \bigl(\operatorname{poly}(n),O(\log n)\bigr)
\]
consists of all functions $f:\{0,1\}^*\to\mathbb{N}$ for which
there is a machine $N$ of this type satisfying
$f(x)=\#\mathrm{acc}_N(x)$ for every input $x$. Every computation
path of $N$ runs in polynomial time and uses $O(\log n)$ work
space; the auxiliary stack is not counted toward the space bound.
Computation paths are distinguished by their sequences of
nondeterministic choices.
\end{definition}

\begin{theorem}[Vinay~\cite{Vinay1991}]
\label{thm:vinay-counting-auxpdm}
Let
\[
 f\in
 \#\mathsf{NAuxPDM}\text{-}\mathsf{TISP}
 \bigl(\operatorname{poly}(n),O(\log n)\bigr).
\]
There is a $\Logspace$-uniform family $(C_n)_{n\ge1}$ of
polynomial-size positive semi-unbounded arithmetic circuits of
depth $O(\log n)$ such that
\[
 C_n(x)=f(x)
 \qquad\text{for every }x\in\{0,1\}^n.
\]
In particular, $f\in\Logspace\text{-}\#\SAC^1$.
\end{theorem}

The next lemma converts the acceptance probability of a
probabilistic auxiliary pushdown machine into an exact counting
problem. The same representation is used again in
Section~\ref{sec:error-reduction}.

\begin{lemma}[Exact Counting Representation]
\label{lem:auxpdm-counting}
Let $M$ be a polynomial-time probabilistic auxiliary pushdown
machine with $O(\log n)$ work space and a read-once random tape,
and let $p_M(x)$ denote its acceptance probability. There is a
logspace-computable polynomial bound $T(n)$ such that the function
\[
 F_M(x)
 =
 \#\{r\in\{0,1\}^{T(n)}:M(x;r)\text{ accepts}\}
\]
belongs to $\Logspace\text{-}\#\SAC^1$ and satisfies
\[
 p_M(x)=2^{-T(n)}F_M(x).
\]
Here the machine ignores any unused random bits.
\end{lemma}

\begin{proof}
Choose $T(n)$ to bound the number of random bits read by $M$. After
the original computation halts, retain its decision and read and
ignore the remaining bits; an $O(\log n)$-bit counter suffices for
this padding. The padded machine still runs in polynomial time,
uses $O(\log n)$ work space, and has the same acceptance
probability.

Replace each random-bit choice by a nondeterministic binary choice,
counting the two choices separately. Since the random tape is read
once, every string in $\{0,1\}^{T(n)}$ determines exactly one
computation path. Hence the resulting nondeterministic auxiliary
pushdown machine has exactly $F_M(x)$ accepting paths, and
\Cref{thm:vinay-counting-auxpdm} gives
$F_M\in\Logspace\text{-}\#\SAC^1$. The identity
$p_M(x)=2^{-T(n)}F_M(x)$ follows because the random strings are
uniformly distributed.
\end{proof}

\begin{proof}[Proof of \Cref{thm:intro-auxpdm-approximation}]
Let $M$ be the probabilistic auxiliary pushdown machine in the
theorem. By \Cref{lem:auxpdm-counting}, there are a
logspace-computable polynomial bound $T(n)$ and a function
$F_M\in\Logspace\text{-}\#\SAC^1$ such that
$p_M(x)=2^{-T(n)}F_M(x)$.

For every logspace-computable integer-valued function $t$ with
$1\le t(n)\le\operatorname{polylog}n$,
\Cref{thm:intro-high-precision} with $K(n)=\Theta(\log n)$
gives a polynomial-size, $\Logspace$-uniform $\AC^1$ circuit family
whose output $\widetilde F_M(x)$ satisfies
\[
 (1-2^{-t(n)})F_M(x)
 \le \widetilde F_M(x)
 \le F_M(x)
\]
and has $t(n)+3$ significand bits and an $O(\log n)$-bit exponent.
Setting $\widetilde p_M(x)=2^{-T(n)}\widetilde F_M(x)$ only
subtracts $T(n)$ from the exponent, which is computable by
$\Logspace$-uniform $\AC^1$ circuits and preserves the
significand and the $O(\log n)$-bit exponent bound. Hence
\[
 (1-2^{-t(n)})p_M(x)
 \le \widetilde p_M(x)
 \le p_M(x),
\]
with the required output representation.
\end{proof}

The acceptance-probability approximation yields the following
bounded-error containment, which restates
\Cref{cor:intro-auxpdm-containment}.

\begin{theorem}
\label{thm:bpauxpdm-ac1}
\[
 \mathrm{BPAuxPDM}\text{-}\mathrm{TISP}
 \bigl(\operatorname{poly}(n),O(\log n)\bigr)
 \subseteq \Logspace\text{-}\AC^1.
\]
\end{theorem}

\begin{proof}
Let $L$ be recognized by a probabilistic auxiliary pushdown machine
$M$ with the stated resource bounds. Apply
\Cref{thm:intro-auxpdm-approximation} with $t=3$ to obtain
\[
 \frac78 p_M(x)\le \widetilde p_M(x)\le p_M(x).
\]
For $x\in L$,
$\widetilde p_M(x)\ge \frac78\cdot\frac23=\frac7{12}>\frac12$,
whereas for $x\notin L$,
$\widetilde p_M(x)\le\frac13<\frac12$.
Comparing the floating-point output with $1/2$ is computable by
$\Logspace$-uniform constant-depth Boolean circuits. Hence
$L\in\Logspace\text{-}\AC^1$.
\end{proof}
\section{\texorpdfstring{$\AC^1$}{AC1} Simulation of Probabilistic Logspace Machines with Bounded Random-Tape Visits}
\label{sec:local_read_k}

\newcommand{\In}{\mathsf{In}}
\newcommand{\Out}{\mathsf{Out}}

We now turn to another application of our multiplicative approximation theorem: probabilistic logspace computation with a locally read-$k$ random tape. We show that the number of accepting random words admits an exact $\Logspace$-$\#\SAC^1$ representation, which in turn yields a $\Logspace$-$\AC^1$ simulation for probabilistic logspace computations with a locally read-$k$ random tape.

\subsection{Model and Theorems}
\label{subsec:local-read-k-model}

We first formalize the locally read-$k$ random-tape model and the corresponding complexity class, and introduce the counting quantities used in the reduction.

\begin{definition}[Locally Read-$k$ Machine]
\label{def:local-random-tape}
A \emph{locally read-$k$ machine} $M$ is a Turing machine equipped with a read-only random tape, such that once the random tape contents are fixed, its computation is deterministic. It satisfies the following conditions:
\begin{itemize}
    \item \textbf{Ordinary tapes.}
    The machine has a two-way read-only input tape and an $O(\log n)$ read--write work tape, and it halts within polynomial time on every input and every random word.

    \item \textbf{Local random tape.}
    The random tape has the form
    \[
    \vdash\, r_1r_2\cdots r_{R(n)}\,\dashv,
    \]
    where $R(n)$ is polynomially bounded and logspace-computable, and $r\in\{0,1\}^{R(n)}$ is chosen uniformly.
    We number the tape positions by $0,1,\ldots,R(n),R(n)+1$,
    with positions $1,\ldots,R(n)$ containing the random bits and positions
    $0$ and $R(n)+1$ containing the fixed endmarkers. When $R(n)\ge1$, the random head starts at position $1$.

    \item \textbf{Bounded visits.}
    At each step, the random head stays at its current position or moves to an adjacent cell.
    A move of the random head from a real cell $i\in\{1,\ldots,R(n)\}$ to $i-1$ or $i+1$ is called a
    \emph{consultation} of cell $i$. A move from an endmarker must move inward to the unique adjacent real cell. The machine satisfies that every real random-tape cell is \emph{consulted at most $k$ times} on every input and every random word.
\end{itemize}
\end{definition}

\begin{definition}[$\LocalRead_k$-$\BPL$]
\label{def:local-read-k-bpl}
The class $\LocalRead_k$-$\BPL$ consists of all languages $L$ for which there exists a locally read-$k$ machine $M$ such that, for every input $x$,
\begin{align*}
    x\in L & \implies \Pr_r[M(x;r)\text{ accepts}]\ge \frac23,\\
    x\notin L & \implies \Pr_r[M(x;r)\text{ accepts}]\le \frac13.
\end{align*}
\end{definition}

Fix a locally read-$k$ machine $M$. Let $T(n)$ be a fixed logspace-computable polynomial upper bound on its running time. Define
\[
F_M(x)=\#\left\{r\in\{0,1\}^{R(n)}:M(x;r)\text{ accepts}\right\},
\qquad
p_M(x)=2^{-R(n)}F_M(x).
\]

The proof proceeds through a parsimonious reduction to $\#\Logspace$. We therefore recall this class.

\begin{definition}[Counting Logspace $\#\Logspace$]
\label{def:sharp-logspace}
A function $f:\{0,1\}^*\to\mathbb{N}$ belongs to $\#\Logspace$ if there is a fixed nondeterministic Turing machine $V$ such that, on every input $x\in\{0,1\}^n$, the machine $V$ uses $O(\log n)$ work space, every computation path halts within $n^{O(1)}$ steps, and the number of accepting computation paths of $V$ on $x$ is exactly $f(x)$. We take nondeterministic choices to be named binary choices: choosing $0$ and choosing $1$ are distinct computation steps even if the resulting configurations later coincide. Deterministic transitions introduce no additional multiplicity.
\end{definition}

Our main structural statement is that bounded local access to the random tape can be encoded parsimoniously in $\#\Logspace$. The standard layered-configuration construction then yields a positive semi-unbounded arithmetic circuit for the same counting function.

\begin{theorem}
\label{thm:local-read-k}
For every fixed $k$, the function $F_M$ belongs to $\#\Logspace$. Moreover, $F_M$ is computed exactly by a $\Logspace$-uniform $\#\SAC^1$ circuit family.
\end{theorem}

Before formally proving \cref{thm:local-read-k}, we use it first. Combining \Cref{thm:local-read-k} with \Cref{thm:approx-sharp-sac} immediately yields the corresponding bounded-error containment.

\begin{theorem}
\label{thm:local-read-k-bpl}
For every fixed $k$,
\[
\LocalRead_k\text{-}\BPL \subseteq \Logspace\text{-}\AC^1.
\]
\end{theorem}

\begin{proof}
Let $M$ recognize a language $L$ in $\LocalRead_k$-$\BPL$. By \Cref{thm:local-read-k}, $F_M$ is computed by a $\Logspace$-uniform $\#\SAC^1$ circuit family. Applying \Cref{thm:approx-sharp-sac} with $t=3$ and dividing by $2^{R(n)}$ gives $\widetilde p_M$ with
\[
\frac78 p_M(x)\le \widetilde p_M(x)\le p_M(x).
\]
Thus $\widetilde p_M(x)>1/2$ on yes-instances and $\widetilde p_M(x)<1/2$ on no-instances. Comparing the output with $1/2$ only needs polynomial-size, $\Logspace$-uniform constant-depth Boolean $\AC$ circuits, and hence $L\in\Logspace$-$\AC^1$.
\end{proof}

We now prove \Cref{thm:local-read-k} for the remaining of this section. 
A key step here is that we can encode each computation using some \(O(k)\) records per random-tape cell, and prove that we can use some local checks between neighboring records to guarantee that the records describe a unique accepting run.

\subsection{Episodes and Local Records}

We first add a deterministic step counter to the work tape. The counter is initialized to zero and incremented after every step of $M$, and therefore uses $O(\log T(n))=O(\log n)$ bits. Its only purpose is to ensure that the local simulations defined below halt within polynomial time, even when started from a guessed surface state.

\begin{definition}[Surface State]
\label{def:surface-state}
A \emph{surface state} $q$ records the finite control (state of $M$), the complete work-tape contents and work-tape head position, the ordinary-input head position, and the step counter. It does not record the ordinary input, the random tape, or the random-head position.

Let $Q_n$ be the set of surface states and let $q_0(n)$ be the initial surface state. Every surface state has a canonical encoding of length $s(n)=O(\log n)$, and hence
\[
|Q_n|\le 2^{s(n)}=n^{O(1)}.
\]
\end{definition}

The purpose of a surface state is to describe everything needed to continue the deterministic computation once the current random-tape position and the symbol stored there are specified.

\begin{definition}[Episode Transition]
\label{def:episode-transition}
Fix a random-tape position $i$, the symbol $r$ stored at position $i$, and a surface state $q\in Q_n$. The \emph{episode transition}
\[
\Delta_{x,i,r}(q)\in
\bigl(\{-1,+1\}\times Q_n\bigr)\cup
\{\mathsf{acc},\mathsf{rej}\}
\]
is defined by simulating $M$ from surface state $q$ with the random head at position $i$, until the first time that either the machine halts or the random head moves.

If the machine halts, $\Delta_{x,i,r}(q)$ returns $\mathsf{acc}$ or $\mathsf{rej}$. If the random head moves by direction $d\in\{-1,+1\}$ and the resulting surface state is $q'$, then
\[
\Delta_{x,i,r}(q)=(d,q').
\]
If the simulation attempts to exceed the prescribed time or work-space bound, or to move a head outside its permitted tape region, we leave \(\Delta_{x,i,r}(q)\) undefined.
\end{definition}

The map $\Delta$ is computable in deterministic $O(\log n)$ space and polynomial time. The key point is that, before the random head moves (to left or to right), all intervening steps are deterministic and the random head remains at the same tape position. Thus the entire segment depends only on $(x,i,r,q)$.
We call such a segment an \emph{episode}. More precisely, in an actual run, an episode starts at the initial configuration or immediately after a random-head move, and ends at the next random-head move or when the machine halts. Therefore the random-head movements partition every run uniquely into episodes.

Suppose that the $h$-th episode starts with the random head at position $i_h$ and surface state $q_h$. Then an actual run determines a unique sequence
\[
(i_0,q_0),(i_1,q_1),\ldots,(i_m,q_m),
\qquad i_0=1,
\]
such that, for every $0\le h<m$,
\[
\Delta_{x,i_h,r_{i_h}}(q_h)
=
(i_{h+1}-i_h,q_{h+1}),
\]
and the final episode has transition outcome $\mathsf{acc}$ or $\mathsf{rej}$.

We emphasize that the episode index $h$ is different from the machine step counter stored in $q_h$: the latter counts all machine steps, whereas $h$ counts only the successive episodes (Left/Right moves of the random tape head).  Since every nonterminal episode contains a random-head move and hence at least one machine step, we have $m\le T(n)$. Let $H:=T(n)+1$ be a convenient uniform upper bound on episode indices.

\begin{lemma}
\label{lem:bounded-episodes}
Every real random-tape cell appears in at most $k+1$ episodes of an actual run, and each endmarker (of the random tape) appears in at most $k$ episodes. Hence for every random tape cell, $K:=k+1$ is a uniform bound on the number of episodes that this cell can occur.
\end{lemma}

\begin{proof}
Every nonterminal episode at a real cell ends when the random head leaves that cell, and this departure is a consultation of the cell. Hence there are at most $k$ such episodes. There may additionally be one terminal episode that halts before another consultation, giving the bound $k+1$.

An episode at an endmarker must begin when the random head enters that endmarker from the adjacent real cell. Each such entrance is caused by a consultation of the adjacent real cell, so there are at most $k$ such episodes.
\end{proof}

The bounded-visit condition therefore implies that the complete interaction history of a single random-tape position can be represented by only constantly many episode records.

\begin{definition}[Local Table]
\label{def:local-table}
For a tape position $i$, a \emph{local table} is a list
\[
S_i=
\bigl(
(h_{i,1},q_{i,1},\alpha_{i,1}),
\ldots,
(h_{i,\ell_i},q_{i,\ell_i},\alpha_{i,\ell_i})
\bigr),
\qquad
0\le \ell_i\le K,
\]
where $0\le h_{i,1}<\cdots<h_{i,\ell_i}\le H$, $q_{i,j}\in Q_n$, and $\alpha_{i,j}\in\{\mathsf{L},\mathsf{R},\mathsf{Start}\}$.

An entry $(h,q,\alpha)$ represents a proposed episode that starts at tape position $i$, at episode index $h$, and in surface state $q$. The label $\alpha=\mathsf{L}$ or $\mathsf{R}$ records whether the random head arrived from the left or right neighboring position, while $\alpha=\mathsf{Start}$ denotes the initial episode.
\end{definition}

\begin{definition}[Local Validity]
For the symbol $r_i$ stored at position $i$, consider an entry $(h,q,\alpha)$ of $S_i$. We call it a \emph{start entry} if $\alpha=\mathsf{Start}$, and a \emph{terminal entry} if $\Delta_{x,i,r_i}(q)=\mathsf{acc}$. An entry whose transition has the form $\Delta_{x,i,r_i}(q)=(d,q')$,
is called \emph{nonterminal}.

We say that $S_i$ is \emph{locally valid} if every entry describes a legal episode at position $i$. Concretely:
\begin{itemize}
    \item A start entry, if present, must be $(0,q_0,\mathsf{Start})$ at position $1$, and every other entry has positive episode index.

    \item The boundary tables contain no impossible arrivals: $S_0$ has no entry with label $\mathsf{L}$, and $S_{R(n)+1}$ has no entry with label $\mathsf{R}$.

    \item For every entry $(h,q,\alpha)$, the transition $\Delta_{x,i,r_i}(q)$ is defined and is not $\mathsf{rej}$, and every nonterminal entry satisfies $h<H$.
\end{itemize}
\end{definition}

A local table describes the complete proposed episode history of one specific tape position. Local validity concerns only the behavior within that position; what remains is to match departures and arrivals across adjacent positions.
We fix a canonical fixed-length encoding of a local table using $K$ slots, with unused slots padded by zeros. Since $K=O(1)$, each table has an $O(\log n)$-bit encoding.

\subsection{Local Compatibility and Accepting Certificates}

For a local table $S_i$, define the multisets
\[
\In_{\mathsf{L}}(S_i)=\{(h,q):(h,q,\mathsf{L})\in S_i\},
\quad
\Out_{\mathsf{L}}(S_i)=\{(h+1,q'):(h,q,\alpha)\in S_i,\ \Delta_{x,i,r_i}(q)=(-1,q')\};
\]
\[
\In_{\mathsf{R}}(S_i)=\{(h,q):(h,q,\mathsf{R})\in S_i\},
\quad
\Out_{\mathsf{R}}(S_i)=\{(h+1,q'):(h,q,\alpha)\in S_i,\ \Delta_{x,i,r_i}(q)=(+1,q')\}.
\]
Thus $\In_{\mathsf{L}}(S_i)$ and $\In_{\mathsf{R}}(S_i)$ record the episodes entering position $i$ from its left and right neighbors, while $\Out_{\mathsf{L}}(S_i)$ and $\Out_{\mathsf{R}}(S_i)$ record the episodes leaving position $i$ in the corresponding directions.

We call two adjacent tables $S_i$ and $S_{i+1}$ \emph{compatible} if
\[
\Out_{\mathsf{R}}(S_i)=\In_{\mathsf{L}}(S_{i+1})
\qquad\text{and}\qquad
\Out_{\mathsf{L}}(S_{i+1})=\In_{\mathsf{R}}(S_i).
\]
In other words, every episode leaving one position is recorded exactly once as the corresponding arrival at each of its neighbors, with the same timestamp and surface state.

\begin{definition}[Accepting Certificate]
\label{def:accepting-certificate}
Fix a random word $r\in\{0,1\}^{R(n)}$ and the input $x \in \{0, 1\}^n$, an \emph{accepting certificate} for $(x,r)$ is a tuple 
\[
\mathcal{S}=(S_0,S_1,\ldots,S_{R(n)+1})
\]
of local tables
such that:
\begin{itemize}
    \item Every $S_i$ is locally valid for the symbol at position $i$.
    \item Every adjacent pair $S_i,S_{i+1}$ is compatible.
    \item Across all tables there is exactly one start entry and exactly one terminal entry.
\end{itemize}
\end{definition}

The definition separates the two ingredients of the certificate: local validity checks each episode in isolation, while compatibility identifies the departure of one episode with the arrival of the next. The remaining point is to verify that these local conditions describe one global computation rather than several disconnected components.

\begin{lemma}
\label{lem:uniqueness}
Fix $x$ and $r$. If $M(x;r)$ accepts, then $(x,r)$ has exactly one accepting certificate; if $M(x;r)$ rejects, then it has none.
\end{lemma}

\begin{proof}
An accepting run uniquely determines a tuple of local tables by recording, at each tape position, the episode index, surface state, and arrival side of every episode beginning there. Each table is locally valid by construction, and every adjacent pair is compatible because each random-head move is recorded once as a departure and once as the corresponding arrival. The run has a unique initial episode and a unique accepting terminal episode, so the resulting tuple has exactly one start entry and exactly one terminal entry. Finally, by \Cref{lem:bounded-episodes}, each table contains at most $K$ entries. Hence the resulting tuple is an accepting certificate.

Conversely, let $\mathcal{S}$ be an accepting certificate. Connect each nonterminal entry $(h,q,\alpha)$ at position $i$ to the unique compatible entry at position $i+d$ with index $h+1$ and state $q'$, where $\Delta_{x,i,r_i}(q)=(d,q')$. Compatibility gives every nonterminal entry a unique successor and every nonstart entry a unique predecessor. Since episode indices increase by one along every edge, there are no directed cycles. With a unique start and a unique terminal entry, all entries therefore form a single directed path from the start to the terminal, and this path is exactly an accepting run of $M(x;r)$. Thus an accepting certificate exists exactly when $M(x;r)$ accepts, and it is then unique.
\end{proof}

\subsection{Exact Counting in Logspace}

We now turn the local description into a parsimonious counting machine.
Recall the definition of function $F_M$ from \cref{def:local-read-k-bpl}, we show the following property for $F_M$.
\begin{lemma}
\label{lem:local-read-k-sharpL}
There is a polynomial-time nondeterministic logspace machine $V$ such that, for every input $x$, the number of accepting computation paths of $V$ on $x$ is exactly $F_M(x)$. Consequently,
\[
F_M\in\#\Logspace.
\]
\end{lemma}

\begin{proof}
The machine $V$ scans the tape positions $i=0,1,\ldots,R(n)+1$ from left to right. At each real cell $i$, it guesses the random bit $r_i$ by one named binary choice; at the two endmarkers, the symbol is fixed. It then guesses the canonical fixed-length encoding of the local table $S_i$, checks that $S_i$ is locally valid, and computes $\In_{\mathsf L}(S_i),\In_{\mathsf R}(S_i),\Out_{\mathsf L}(S_i),\Out_{\mathsf R}(S_i)$. For $i\ge 1$, it checks compatibility with the preceding table by verifying $\Out_{\mathsf R}(S_{i-1})=\In_{\mathsf L}(S_i)$ and $\Out_{\mathsf L}(S_i)=\In_{\mathsf R}(S_{i-1})$. Afterward it keeps only $\bigl(\Out_{\mathsf R}(S_i),\In_{\mathsf R}(S_i)\bigr)$ for the next position.

The machine also records whether a start entry and a terminal entry have already been seen. If a second start entry or a second terminal entry appears, it rejects. After the final table is processed, $V$ accepts exactly when both a start entry and a terminal entry have been seen. Thus an accepting computation of $V$ corresponds exactly to a pair $(r,\mathcal S)$ where $\mathcal S$ is an accepting certificate for $(x,r)$.

By \Cref{lem:bounded-episodes}, each local table and each boundary multiset contains at most $K = O(1)$ records, each using $O(\log n)$ bits, so $V$ uses $O(\log n)$ work space. Checking local validity requires only constantly many simulations of $\Delta$, each of which runs for at most $T(n)$ steps. Since there are only $R(n)+2$ tape positions, every computation path of $V$ halts in polynomial time.

All nondeterminism is used only to guess the random word and the canonical table encodings; once these choices are fixed, the verification is deterministic. Hence each pair $(r,\mathcal S)$ determines exactly one computation path of $V$. By \Cref{lem:uniqueness}, for each random word $r$, there is exactly one accepting certificate if $M(x;r)$ accepts, and none otherwise. Therefore the number of accepting computation paths of $V$ is exactly $F_M(x)$, and hence $F_M\in\#\Logspace$.
\end{proof}

This proves the first assertion of \Cref{thm:local-read-k}. The case $R(n)=0$ is immediate: there is a single empty random word, and the deterministic polynomial-time logspace computation of $M$ determines whether the count is zero or one.

\subsection{From $\#\Logspace$ to $\Logspace$-$\#\SAC^1$}

We use the standard inclusion
\[
\#\Logspace \subseteq \Logspace\text{-}\#\SAC^1,
\]
which follows from the layered configuration graph of a polynomial-time logspace counting machine.

\begin{lemma}[$\#\Logspace \subseteq \Logspace$-$\#\SAC^1$]
\label{lem:sharpL-to-sharpSAC}
Every function in $\#\Logspace$ is computed by a $\Logspace$-uniform $\#\SAC^1$ circuit family.
\end{lemma}

\begin{proof}
Let $V$ be a polynomial-time nondeterministic logspace machine computing $f\in\#\Logspace$, and pad all computation paths to a common polynomial length $L$. Since $V$ is logspace and has only $n^{O(1)}$ configurations, its layered configuration graph has polynomial width. For each layer $j$, let $B_j(x)$ be the matrix whose $(u,v)$ entry is the number of named one-step transitions from configuration $u$ to configuration $v$. Then
\[
f(x)=e_{\mathrm{start}}^{\mathsf T}B_0(x)B_1(x)\cdots B_{L-1}(x)e_{\mathrm{acc}}.
\]
Here $e_{\mathrm{acc}}$ is the indicator vector of the accepting configurations in the final layer.
Each matrix entry is a sum of constantly many literals or constants, and a balanced product tree computes the matrix product with polynomial size and depth $O(\log L)=O(\log n)$ using unbounded addition and binary multiplication. The construction is $\Logspace$-uniform since $V$ is logspace.
\end{proof}

By \Cref{lem:local-read-k-sharpL} and \Cref{lem:sharpL-to-sharpSAC}, the counting function $F_M$ lies in $\Logspace$-$\#\SAC^1$, proving \Cref{thm:local-read-k}.

\section{High-Precision Approximation via Error Reduction}
\label{sec:error-reduction}

\Cref{thm:approx-sharp-sac} gives multiplicative approximations to positive arithmetic circuits of depth $K$ in Boolean depth $O(K+t)$. We improve the dependence on $t$ by simultaneously refining the gate values and an approximate inverse of the global Jacobian.

For convenience, put
\[
d(n,K,t)=K+\frac{\log n}{\log\log n}\log\left(2+\frac{t}{K}\right).
\]
For $K=O(\log n)$ we write $d(n,t)=d(n,\lceil\log n\rceil,t)$.

We restate \Cref{thm:intro-high-precision}
using the notation above.

\begin{theorem}[High-Precision Approximation of Positive Arithmetic Circuits]
\label{thm:er-high-precision}
Let $K(n),t(n)\ge1$ be logspace-computable, polynomially bounded integer-valued functions. Let $(F_n:\{0,1\}^n\to\mathbb{N}_{\ge0})_{n\ge1}$ be a function family computed by a polynomial-size, $\Logspace$-uniform family $(G_n)_{n\ge1}$ of positive semi-unbounded arithmetic circuits of depth at most $K(n)$. Then there is a polynomial-size, $\Logspace$-uniform Boolean $\AC$ circuit family of depth $O(d(n,K(n),t(n)))$ that, on every input $x\in\{0,1\}^n$, outputs a nonnegative $(t(n)+3,O(K(n)+\log n))$-number $\widetilde F_n(x)$ satisfying
\[
(1-2^{-t(n)})F_n(x)\le\widetilde F_n(x)\le F_n(x).
\]
In particular, when $K(n)=O(\log n)$, every fixed polylogarithmic precision $t(n)=(\log n)^{O(1)}$ is attainable in $\Logspace$-uniform $\AC^1$.
\end{theorem}

To show the theorem, we first use \Cref{thm:approx-sharp-sac}  to approximate all gate values and construct a positive quadratic system whose coefficients have polynomial precision and \(O(\log n)\)-bit exponents. Then we reduce the error by simultaneously refining the gate-value approximations and an approximate inverse of the Jacobian.

\subsection{Preparing the System and the Initialization for Error Reduction}
\label{subsec:er-init}

We first put the arithmetic circuit into a layered form of polynomial width and depth \(K\), where \(K\) is a logspace-computable upper bound on the original depth. To preserve logspace uniformity without computing individual gate depths, we create a copy of every gate at each layer. At layer \(0\), source gates hold their prescribed values and all other gates are initialized to zero. At each subsequent layer, source gates retain their values, while every other gate applies its operation to the copies of its predecessors in the preceding layer. By layer \(K\), every gate has its correct value. This construction increases the circuit size by a factor of at most \(O(K)\). Let \(N\ge 2\) be a logspace-computable polynomial upper bound on the number of gates and wires in the resulting circuit.

Fix a Boolean input $x$. Let $x_0,\ldots,x_K$ be the exact layer-value vectors, where $x_i=\Phi_i(x_{i-1})$ and each coordinate of $\Phi_i$ is a constant, a copy, a positive sum, or a binary product. Write $X=(x_0,\ldots,x_K)$ and define the global map by
\[
\Phi(z)_0=x_0,\qquad \Phi(z)_i=\Phi_i(z_{i-1})\quad(1\le i\le K).
\]
Thus $X=\Phi(X)$. Since every coordinate has degree at most two and nonnegative coefficients, we may write $\Phi(z)=c+Lz+Q(z)$, where $L$ is linear and $Q$ is homogeneous quadratic, and put $J(z)=D\Phi(z)$. 
Since each layer depends only on the preceding layer, \(J(z)\) is strictly block lower triangular and hence satisfies \(J(z)^{K+1}=0\).

Set 
\begin{equation}\label{eq:initparam}
    \tau=6K+20, \eta=2^{-\tau}, \mbox{ and } \rho=\eta/4.
\end{equation} 
The layerwise construction of \Cref{thm:approx-sharp-sac} computes, in Boolean depth $O(K+\tau)=O(K)$, vectors $\bar b_0=x_0,\bar b_1,\ldots,\bar b_K$ such that
\[
(1-\rho)x_i\le\bar b_i\le x_i\qquad(1\le i\le K).
\]
Each entry of \(\bar b_i\) is a \((\tau+4,O(K+\log n))\)-number. When \(K\) is polynomially large, the exact gate values may require exponentially many bits in binary. To control their magnitudes, we rescale these values by powers of two determined by the exponents of their approximations.

We first rescale the system so that its nonzero gate values have bounded magnitude, and then simplify it using the known zero coordinates. 
For each nonzero coordinate $v$ outside the input block, write $\bar b_v=M_v2^{e_v-\ell_v}$ and put $\sigma_v=2^{e_v}$. Set $\sigma_v=1$ on zero coordinates and on the input block, and define
\[
S=\operatorname{diag}(\sigma_v),\qquad Z=S^{-1}X,\qquad \Psi(z)=S^{-1}\Phi(Sz).
\]
Thus $Z=\Psi(Z)$ since $X=\Phi(X)$. The zero coordinates of $Z$ are known from the zero flags of $\bar b$ (since \(\bar b\) approximates the exact vector \(X\), and \(Z\) is obtained by rescaling that same vector \(X\)). For each non-input coordinate of $\Psi$, delete monomials containing a variable whose coordinate in $Z$ is zero, and set the entire row to zero if its coordinate in $Z$ is zero. This leaves the fixed point $Z$ unchanged. Keep parallel-wire occurrences as separate monomials.
Put
\[
P=K+\lceil\log N\rceil+\max\{t,\tau\}+10.
\]
Replace the coefficient $a$ of each remaining monomial by $\widehat a=2^{-P}\lfloor2^Pa\rfloor$, and denote the resulting layered positive quadratic map by $\widehat\Phi$, keeping its input block equal to $x_0$. Let $\widehat X=(\widehat x_0,\ldots,\widehat x_K)$ be its exact layer-value vector. Finally, define $\widehat b_0=x_0$ and $\widehat b_v=(1-\rho)\bar b_v/\sigma_v$ outside the input block. 

The next lemma shows that the above transformation—consisting of rescaling, simplification, and coefficient truncation—introduces only a small multiplicative error.
\begin{lemma}
\label{lem:er-scaling}
With the above definitions, put $\epsilon=8N(2^K-1)2^{-P}<2^{-\max\{t,\tau\}-7}$. The scaled exact vector $Z$ and the exact layer-value vector $\widehat X$ satisfy
\[
(1-\epsilon)Z\le\widehat X\le Z,\qquad (1-2^{-\tau})\widehat x_i\le\widehat b_i\le\widehat x_i\quad(0\le i\le K).
\]
Every nonzero coordinate of $Z$ belongs to $[1/2,2]$. Each nonzero coefficient of $\widehat\Phi$ is a $(1,O(\log n))$-number, and its value is a multiple of $2^{-P}$ bounded by $8$. Each entry of $\widehat b_i$ is a $(2\tau+6,O(\log n))$-number.
\end{lemma}

\begin{proof}
By the approximation guarantee, $\bar b_v=0$ if and only if $X_v=0$. For each nonzero coordinate outside the input block, the floating-point representation gives
\[
\frac12\le\frac{\bar b_v}{\sigma_v}<1,\qquad 1-\rho\le\frac{\bar b_v}{X_v}\le1.
\]
Consequently, $1/2\le Z_v<1/(1-\rho)<2$. The input block is unchanged, so its nonzero coordinates also lie in $[1/2,2]$. Since each $\sigma_v$ is a power of two, the scaling matrix $S$ is represented implicitly by the exponent fields $e_v$; all scaled coefficients can be computed from these exponents without explicitly constructing $S$.

Each retained coefficient $a$ of $\Psi$ is a positive power of two. At $Z$, the variable part of its monomial has value at least $1/4$ (and a constant monomial has variable part $1$), while its contribution is bounded above by the corresponding output $Z_v\le2$. Hence $a\le8$. Since $a$ is a power of two, replacing it by $\widehat a$ either preserves it exactly or sets it to zero. Every term removed by coefficient truncation contributes less than $4\cdot2^{-P}$ at $Z$. There are at most $N$ monomial occurrences per coordinate, so with $\delta=8N2^{-P}$ we obtain
\[
(1-\delta)\Psi(Z)\le\widehat\Phi(Z)\le\Psi(Z)=Z.
\]
The inequality includes zero coordinates, whose rows have been set to zero. Since the system is positive and layered, $\widehat X\le Z$. If $\widehat x_{i-1}\ge(1-\epsilon_{i-1})Z_{i-1}$, positivity and the degree-two bound imply
\[
\widehat x_i\ge(1-\epsilon_{i-1})^2(1-\delta)Z_i\ge(1-2\epsilon_{i-1}-\delta)Z_i.
\]
Starting from $\epsilon_0=0$ gives $\epsilon_i\le(2^i-1)\delta$, and hence $(1-\epsilon)Z\le\widehat X\le Z$.

To verify the approximation guarantee for $\widehat b$, consider the non-input coordinates. The definitions and the original guarantee give
\[
(1-\rho)^2Z_v\le\widehat b_v\le(1-\rho)Z_v.
\]
As $\epsilon\le\rho$, we have $(1-\rho)Z_v\le(1-\epsilon)Z_v\le\widehat X_v$, while $(1-\rho)^2\ge1-\eta$ and $\widehat X_v\le Z_v$. Thus $(1-\eta)\widehat X_v\le\widehat b_v\le\widehat X_v$. The input block satisfies the same conclusion exactly, since $\widehat b_0=\widehat x_0=x_0$. Finally, $\bar b_v/\sigma_v=M_v2^{-\ell_v}$ with $\ell_v\le\tau+4$, and $1-\rho=(2^{\tau+2}-1)2^{-(\tau+2)}$. Hence the precision bound of each $\widehat b_v$ is at most $2\tau+6$, and its exponent requires $O(\log n)$ bits. Each retained coefficient of $\widehat\Phi$ is a power of two in $[2^{-P},8]$, so it is a $(1,O(\log n))$-number.
\end{proof}

From now on, we use $\Phi,X,b_i$ for the scaled and truncated objects $\widehat\Phi,\widehat X,\widehat b_i$, and write $X^{\mathrm{orig}}$ for the original gate-value vector. By \Cref{lem:er-scaling},
\begin{equation}
(1-\epsilon)X^{\mathrm{orig}}\le SX\le X^{\mathrm{orig}},\qquad 0\le \|X\|_\infty\le2.
\label{eq:er-original-values}
\end{equation}
The scaled and truncated map \(\Phi\) retains nonnegative coefficients, degree at most two, and strictly layered dependencies.

To initialize the error reduction, we need approximations with nonnegative residuals, which the bounds \(b_i\le x_i\) alone do not guarantee. The next lemma slightly decreases the layerwise approximations to obtain \(y\le\Phi(y)\), ensuring that \(r=\Phi(y)-y\ge0\). It also constructs an approximate Jacobian inverse \(R\) with nonnegative inverse residual \(E=I-(I-J(y))R\). The resulting bounds on the gate-value error and inverse residual provide the starting estimates for the simultaneous refinement in \cref{subsec:er-refinement}.

\begin{lemma}
\label{lem:er-init}
For the approximations $b_i$ obtained from \Cref{lem:er-scaling}, define
\[
y_0=x_0,\qquad y_i=(1-4^i\eta)b_i\quad(1\le i\le K).
\]
\begin{enumerate}
 \item  Then $y$ is a lower solution of the global system, with
\[
0\le y\le\Phi(y)\le X,\qquad X-y\le\alpha_0X,\qquad \alpha_0=2\cdot4^K\eta.
\]
In particular, the exact residual $r=\Phi(y)-y$ is nonnegative.

    \item Put $J=J(y)=D\Phi(y)$ and
\[
G_\eta(y)=(I-(1-\eta)J)^{-1}=\sum_{j=0}^K(1-\eta)^jJ^j.
\]
Let $R$ be the lower approximation to $G_\eta(y)$ given by \Cref{lem:shallow-positive-circuit} at accuracy $\tau$, using a geometric-doubling circuit of arithmetic depth $O(\log(K+2))$. Then
\[
(1-\eta)G_\eta(y)\le R\le G_\eta(y).
\]
Also we have $R\ge0$, and the inverse residual $E=I-(I-J)R$ satisfies
\[
E\ge0,\qquad EX\le\beta_0X,\qquad \beta_0=(K+1)\eta H,\quad H=2^{K+1}-1.
\]

\end{enumerate}
Every entry of $y,r,R,E$ is an $(n^{O(1)},O(\log n))$-number.
\end{lemma}

\begin{proof}
The initial approximation guarantee yields
\[
[1-(4^i+1)\eta]x_i\le y_i\le(1-4^i\eta)x_i.
\]
The essential point is to ensure $y_i\le\Phi_i(y_{i-1})$ at every layer. For a positive polynomial of degree at most two and $0\le\lambda\le1$, we have $\Phi_i(\lambda u)\ge\lambda^2\Phi_i(u)$. Thus for $i\ge2$,
\[
\Phi_i(y_{i-1})\ge[1-(4^{i-1}+1)\eta]^2x_i\ge(1-4^i\eta)x_i\ge y_i.
\]
For $i=1$, the assertion follows from $y_0=x_0$. Monotonicity gives $\Phi(y)\le\Phi(X)=X$ and $y\ge0$ is immediate. Also, $X-y\le(4^K+1)\eta X\le2\cdot4^K\eta X$. 
This proves the first part.

Write $G=G(y)=(I-J(y))^{-1}$ and $G_\eta=G_\eta(y)$. Since $(1-\eta)JG_\eta=G_\eta-I$, the lower approximation guarantee implies
\[
JR\ge(1-\eta)JG_\eta=G_\eta-I,\qquad E=I-R+JR\ge G_\eta-R\ge0.
\]
Moreover, $0\le R\le G_\eta\le G$, while
\[
R\ge(1-\eta)G_\eta=\sum_{j=0}^K(1-\eta)^{j+1}J^j\ge(1-(K+1)\eta)G.
\]
Putting $\Delta=G-R\ge0$, we have $0\le E=(I-J)\Delta\le\Delta$. Since $JX\le2X$ by positivity and the degree-two bound, we have $GX=\sum_{j=0}^KJ^jX\le HX$, and therefore
\[
EX\le\Delta X\le(K+1)\eta GX\le\beta_0X.
\]

To establish the precision and exponent-length bounds, write each $b_i$ exactly as $A_i2^{-D_i}$, where $D_i=O(\tau)$ and $A_i$ has $O(\tau)$ bits by \Cref{lem:er-scaling}. The factors $1-4^i\eta$ preserve an $O(\tau)$ bound on $D_i$. The coefficients of $\Phi$ are multiples of $2^{-P}$, so each entry of $(1-\eta)J$ is an exact multiple of $2^{-D_J}$, where $D_J=O(P+\tau)$, and has value at most $16N$. Every nonzero entry of $G_\eta$ is at least $2^{-KD_J}$ and at most $(K+1)(16N^2)^K$. The approximation $R$ has precision at most $\tau+O(\log(K+2))$ by \Cref{lem:shallow-positive-circuit}, and its nonzero entries lie within a factor of two of those of $G_\eta$. Consequently, the entries of $R$, and hence those of $y,r,E$ and the exact intermediate operations used to initialize them, have a common representation $A2^{-D_0}$ with
\[
D_0=O(K(P+\tau)),\qquad W_0=D_0+O(K\log N),
\]
where $A$ has at most $W_0$ bits. These polynomially bounded, logspace-computable bounds give the asserted $(n^{O(1)},O(\log n))$-representations.
\end{proof}

\subsection{Simultaneous Refinement for Error Reduction}
\label{subsec:er-refinement}
Consider an approximation $y$ for gate values $X$, the approximation bias vector $r = \Phi(y) - y$, a matrix $R(y)$ which is an approximation for $G(y) = (I-J(y))^{-1}$, and $E(y):=I-(I-J(y))R$ which is an inverse residue of $R$.
Assume we start with $(y, r, R, E)$ with the following properties
\begin{equation}
0\le y\le X, \qquad
r=\Phi(y)-y\ge0,\qquad R\ge0,\qquad E=I-(I-J(y))R\ge0.
\label{eq:er-invariants}
\end{equation}

We update the gate values approximation $y$, the approximation bias $r$, the inverse matrix approximation $R$ and the inverse residual $E$ for each round in the following way.
Define the updated gate-value approximation \(y'\),  approximation residue $r'$, inverse matrix approximation $R'$, and inverse matrix residual \(E'\) by the following quantities
\begin{equation}\label{def:y' and E'}
    \begin{aligned}
        s&=Rr, & y'&=y+s,\\
        r'&=Er+Q(s), & B&=E+DQ(s)R,\\
        R'&=R+RB, & E'&=B^2.
\end{aligned}
\end{equation}

\begin{lemma}
\label{lem:er-update}
Suppose  \eqref{eq:er-invariants} hold. 
Then $0\le y\le y'\le\Phi(y')\le X$, and
\[
r'=\Phi(y')-y'\ge0,\qquad R'\ge0,\qquad
E'=I-(I-J(y'))R'\ge0.
\]
\end{lemma}

\begin{proof}
Let $e=X-y$ and $G(y)=(I-J(y))^{-1}$. Since $X=\Phi(X)$ and $\Phi$ has degree at most two, its exact Taylor expansion $X = \Phi(X) = \Phi(y) + J(y) e + Q(e)$ gives
\[
(I-J(y))e=r+Q(e),\qquad e=G(y)(r+Q(e)).
\]
The inverse exists and is nonnegative because $J(y)$ is nonnegative and nilpotent. From $(I-J(y))R=I-E$, we obtain
\[
R=G(y)-G(y)E\le G(y).
\]
Consequently, $0\le s=Rr\le G(y)r\le e$, proving $0\le y\le y'\le X$. Expanding $\Phi$ at $y$ gives the new residual directly:
\[
\Phi(y+s)-(y+s)=r-(I-J(y))s+Q(s)=Er+Q(s)=r'.
\]
Finally, $J(y+s)-J(y)=DQ(s)$, so $(I-J(y'))R=I-B$. With $R'=R(I+B)$, we obtain
\[
(I-J(y'))R'=(I-B)(I+B)=I-B^2.
\]
Thus $E'=B^2=I-(I-J(y'))R'$. Positivity of $r,R,E,Q$ and $DQ(s)$ proves the remaining invariants. The entries of $DQ(s)$ are explicit from the quadratic monomials of $\Phi$: a term $c_{ab}z_az_b$ contributes $c_{ab}s_b$ in column $a$ and $c_{ab}s_a$ in column $b$, while lower-degree terms contribute zero.
\end{proof}

The gate-value error and inverse residual both decrease quadratically, provided that their initial values are sufficiently small.

\begin{lemma}
\label{lem:er-convergence}
 Suppose  \eqref{eq:er-invariants} hold and $X-y\le\alpha X$, $EX\le\beta X$ for nonnegative scalars $\alpha,\beta$. 
The updated gate-value approximation \(y'\) and inverse residual \(E'\) from \eqref{def:y' and E'} satisfy
\[
    X-y'\le\alpha'X,\qquad E'X\le\beta'X,
\]
where 
\[
    \alpha'=H(\alpha^2+\beta\alpha),\qquad \beta'=(\beta+2H\alpha)^2,
\]
and $H:=2^{K+1}-1$.
In particular, if $\zeta=\max\{H^2\alpha,H\beta\}$ and $\zeta'=\max\{H^2\alpha',H\beta'\}$, then $\zeta'\le9\zeta^2$.
\end{lemma}

\begin{proof}
Recall that we write $\Phi(z)=c+Lz+Q(z)$, where $c$ is the constant term,
$L$ is the matrix of linear coefficients, and $Q$ is homogeneous
quadratic. Since $0\le y\le X$, all coefficients are nonnegative,
and $\Phi(X)=X$, we have
\[
    J(y)X\le J(X)X=LX+2Q(X)\le 2X.
\]
Since $G(y)=\sum_{j=0}^KJ(y)^j$,
\[
G(y)X\le\left(\sum_{j=0}^K2^j\right)X=HX.
\]
Put $e=X-y$ and recall $r=\Phi(y)-y$. The quadratic expansion $\Phi(X)=\Phi(y)+J(y)e+Q(e)$ gives $e=r+J(y)e+Q(e)$, hence $0\le r\le e$ and $e=G(y)(r+Q(e))$. 
Also by \eqref{eq:er-invariants}, we have $R=G(y)-G(y)E$.
So
\[
X-y'=e-Rr=G(y)\bigl(Q(e)+Er\bigr)\le H(\alpha^2+\beta\alpha)X,
\]
where we use $Q(e)\le\alpha^2Q(X)\le\alpha^2X$ and $Er\le\alpha EX\le\beta\alpha X$. For the inverse residual $E'$, notice that $s\le e\le\alpha X$ implies $DQ(s)X\le\alpha DQ(X)X=2\alpha Q(X)\le2\alpha X$. Furthermore $RX\le G(y)X\le HX$, and therefore
\[
BX=EX+DQ(s)RX\le(\beta+2H\alpha)X.
\]
Applying the nonnegative matrix $B$ once more gives $E'X=B^2X\le(\beta+2H\alpha)^2X$. Finally, by our definition of $\zeta$, we have $H^2\alpha'\le(1+H^{-1})\zeta^2\le2\zeta^2$ and $H\beta'\le9H^{-1}\zeta^2\le9\zeta^2$, proving the joint bound.
\end{proof}

\begin{lemma}\label{lem:after q rounds}
   Suppose \eqref{eq:er-invariants} holds and
\[
    X-y\le\alpha_0X,\qquad EX\le\beta_0X,
\]
where
\[
    \alpha_0=2\cdot4^K2^{-\tau},\qquad
    \beta_0=(K+1)2^{-\tau}H,\qquad
    H=2^{K+1}-1,
\]
and $\tau$ is as in \eqref{eq:initparam}.

Let
\begin{equation}
\label{eq:er-round-count}
    q=\min\{j\in\mathbb{Z}_{\ge0}:(2K+13)2^j\ge t+4\}.
\end{equation}
Then $q=O(\log(2+t/K))$, and after $q$ rounds of the
updates in \eqref{def:y' and E'}, the resulting vector
$y^{(q)}$ satisfies
\[
    0\le X-y^{(q)}\le2^{-t-4}X.
\]

\end{lemma}
\begin{proof}
For the initial state,  recall that we take $\alpha_0=2\cdot4^K2^{-\tau}$ and $\beta_0=(K+1)2^{-\tau}H$ (in \cref{lem:er-init}). Since $K+1\le2\cdot4^K$, we have $H\beta_0\le H^2\alpha_0$. Using $H<2^{K+1}$, and $\tau=6K+20$ (from \eqref{eq:initparam}), we obtain $\zeta_0\le H^2\alpha_0<2^{4K+3-\tau}=2^{-2K-17}$. Thus $w_j:=9\zeta_j$ satisfies $w_{j+1}\le w_j^2$ and $w_0\le2^{-(2K+13)}$. Inductively,
\[
\alpha_j\le2^{-(2K+13)2^j}.
\]

Let $q$ be the least nonnegative integer with $(2K+13)2^q\ge t+4$. Then
\begin{equation*}
q=O\left(\log\left(2+\frac{t}{K}\right)\right),\qquad 0\le X-y^{(q)}\le2^{-t-4}X.
\end{equation*}

\end{proof}


\subsection{Exact Arithmetic and Circuit Complexity}
\label{subsec:er-complexity}

Every refinement step consists of a constant number of matrix and vector operations. Unlike the earlier approximate evaluations, these operations are exact: their inputs and outputs are $(n^{O(1)},O(\log n))$-numbers, and no rounding is performed before the final output. To analyze the precision bounds, we write each intermediate exactly as $A2^{-D}$, with an integer numerator $A$ and a fixed, logspace-computable parameter $D$.

We use \Cref{lem:er-arithmetic} for exact multiplication and summation and \Cref{lem:elementary-arithmetic} for exact subtraction. The next lemma verifies the required polynomial precision bounds and $O(\log n)$ exponent-length bounds. Hence each exact matrix operation has Boolean depth $O(\log n/\log\log n)$.

\begin{lemma}
\label{lem:er-word-length}



Fix a constant $a>0$, suppose that $t(n)\le n^a$, and let $q  $
be the number of error-reduction rounds specified in \eqref{eq:er-round-count}.
There exist logspace-computable integer-valued functions
$p(n)$ and $b(n)$ satisfying
\[
    p(n)=n^{O(1)},
    \qquad
    b(n)=O(\log n),
\]
such that, for every input $x\in\{0,1\}^n$, every initial tuple $(y,r,R,E)$ constructed in \Cref{lem:er-init}, it holds that all vectors and
matrices computed during these $q$ rounds have their entries being
$(p(n),b(n))$-numbers.

\end{lemma}

\begin{proof}
By \Cref{lem:er-init}, every initial entry can be written as $A2^{-D_0}$ with a common $D_0=n^{O(1)}$ and $A$ of at most $W_0=n^{O(1)}$ bits, where $W_0\ge D_0$. Include the fixed coefficients of $\Phi$ as additional registers whose values are preserved at each round, enlarging $D_0,W_0$ if necessary. Viewed as a polynomial map in these registers and the entries of $(y,r,R,E)$, one refinement step has degree at most eight: $s=Rr$ has degree two, $Q(s)$ degree five, $DQ(s)R$ degree four, $R+RB$ degree at most five, and $B^2$ degree at most eight. 
Since each round contains constantly many operations on polynomial-dimensional arrays, every output has polynomially many monomial occurrences and coefficients of polynomial magnitude.

If the inputs to round $j$ can be written as $A2^{-D_j}$ with at most $W_j$ numerator bits, we may choose
\[
D_{j+1}=8D_j,\qquad W_{j+1}=8W_j+C\lceil\log n\rceil
\]
for a fixed constant $C$. Indeed, a monomial of degree $d\le8$, expressed with the common factor $2^{-D_{j+1}}$, needs at most $dW_j+(8-d)D_j\le8W_j$ numerator bits; summing polynomially many terms adds $O(\log n)$ bits. 

Thus $D_j,W_j\le8^j n^{O(1)}$ for $0\le j\le q$.
Since $q=O(\log(t+2))$ and $t\le n^a$, these bounds remain
polynomial in $n$.

For $0\le j<q$, every intermediate value produced during round
$j$ can be written as $A2^{-D_{j+1}}$, where
$0\le A<2^{W_{j+1}}$.
Removing trailing zeros from the binary representation of $A$
gives the representation in \Cref{def:floating-point}, with
precision at most $W_{j+1}$ and exponent magnitude at most
$D_{j+1}+W_{j+1}$.

Set
\[
    p(n)=W_q,
    \qquad
    b(n)=2+\left\lceil\log_2(D_q+W_q+1)\right\rceil.
\]
Since $D_j$ and $W_j$ are nondecreasing, these bounds apply
to the initial state and every intermediate value in all $q$
rounds. They satisfy $p(n)=n^{O(1)}$ and $b(n)=O(\log n)$.
The initial bounds and $q$ are logspace-computable, and the
recurrences involve only polynomially bounded integers.
Hence $p(n)$ and $b(n)$ are also logspace-computable.
\end{proof}

\begin{proof}[Proof of \Cref{thm:er-high-precision}]
Use the construction underlying \Cref{thm:approx-sharp-sac} at precision $\tau+2=6K+22$ to obtain the vectors $\bar b_i$, simultaneously, with polynomial size and Boolean depth $O(K)$. Apply \Cref{lem:er-scaling} and initialize $y,r,R,E$ as in \Cref{lem:er-init}. The scaling factors are stored only through their exponents. A scaled monomial coefficient is $2^k$, where $k$ is the sum of at most two child exponents minus the output exponent. For its retained numerator over $2^P$, bit $j$, with $0\le j\le P+3$, is one exactly when the monomial survives the zero-coordinate tests and $k=j-P$. These tests use constant-depth Boolean comparisons and arithmetic on the exponent fields. Thus all coefficients are generated in fixed polynomial-length words without expanding the scaling factors. The rescaled approximations $\widehat b_i$ and the factors $1-\rho$ and $1-4^i\eta$ have polynomial precision and $O(\log n)$-bit exponents; the initial residual is computed using exact operations from \Cref{lem:elementary-arithmetic} and \Cref{lem:er-arithmetic}. Compute $J(y)$ exactly using \Cref{lem:elementary-arithmetic} and \Cref{lem:er-arithmetic}, in Boolean depth $O(\log n/\log\log n)$. To initialize $R$, evaluate the positive arithmetic circuit for $G_\eta(y)$ approximately as in \Cref{lem:er-init}. Geometric doubling gives this circuit arithmetic depth $h=O(\log(K+2))$; by \Cref{lem:shallow-positive-circuit}, its lower approximation has Boolean depth
\[
O\left(h\left(1+\frac{\min\{\tau+h,\log n\}}{\log\log n}\right)\right)=O(K).
\]
Here $\tau=6K+20$; the last bound follows by considering $K\le\log n$ and $K>\log n$ separately. 
Compute the initial residuals as
\[
    r=[\Phi(y)-y]_+,\qquad
    E=[I+J(y)R-R]_+,
\]
where $[\cdot]_+$ is applied entrywise.
By \Cref{lem:er-init}, the differences inside the brackets
are nonnegative, so these expressions equal
$\Phi(y)-y$ and $I-(I-J(y))R$, respectively.
They can therefore be computed using a constant number of
exact matrix and vector operations on nonnegative operands,
each of Boolean depth $O(\log n/\log\log n)$ by
\Cref{lem:elementary-arithmetic,lem:er-arithmetic}.
The necessary polynomial precision and $O(\log n)$-bit exponent bounds follow from \Cref{lem:er-init}. Thus the total initialization depth is $O(K+\log n/\log\log n)$.

Take $q$ as defined in \eqref{eq:er-round-count}, and perform
$q$ rounds of the updates in \eqref{def:y' and E'} without
rounding. By \Cref{lem:er-update}, the invariants
\eqref{eq:er-invariants} hold throughout.
By \Cref{lem:after q rounds}, the resulting vector $y^{(q)}$
satisfies $0\le X-y^{(q)}\le2^{-t-4}X$.
Apply \Cref{lem:elementary-arithmetic} to round each coordinate
of $y^{(q)}$ downward to $p=t+3$ significand bits. Define
\[
    \widehat b_i(t)
    =\operatorname{RoundDown}_{p}(y_i^{(q)})
    \quad(0\le i\le K),
    \qquad
    \widehat b(t)
    =(\widehat b_0(t),\ldots,\widehat b_K(t)),
\]
where rounding is applied coordinatewise.
The directed rounding error is at most $2^{1-p}y_i^{(q)}\le2^{-t-2}x_i$. Combining this bound with \eqref{eq:er-original-values} gives
\[
0\le X^{\mathrm{orig}}-S\widehat b(t)\le(\epsilon+2^{-t-4}+2^{-t-2})X^{\mathrm{orig}}<2^{-t}X^{\mathrm{orig}}.
\]

The desired output $\widetilde F_n(x)$ is the designated coordinate of $S\widehat b(t)$ in layer $K$. Zero coordinates remain zero. By \Cref{lem:elementary-arithmetic}, the final rounding to precision $p=t+3$ has constant Boolean depth. Restoring the scale adds the stored exponent of $\sigma_v$, so the final output is a $(t+3,O(K+\log n))$-number.

By \Cref{lem:elementary-arithmetic}, \Cref{lem:er-arithmetic} and \Cref{lem:er-word-length}, every refinement round has polynomial size and Boolean depth $O(\log n/\log\log n)$. Together with the initialization and \eqref{eq:er-round-count}, this gives total depth
\[
O\left(K+\frac{\log n}{\log\log n}\log\left(2+\frac{t}{K}\right)\right)=O(d(n,K,t)).
\]

The $q$ rounds and all their matrices have polynomial size. For uniformity, index each Boolean gate by its stage or round, array type, row and column indices, bit position, and local arithmetic-gate index. These fields use $O(\log n)$ bits, and the original $\Logspace$-uniform arithmetic wiring, the explicit update formulas, and the logspace-computable word-length bounds determine every connection in logarithmic space. The zero-coordinate tests and scaled coefficients are computed as data in fixed circuit locations; they do not change the wiring. This completes the proof.
\end{proof}

\subsection{Consequences for Probabilistic Computation}
\label{subsec:er-machines}

\Cref{lem:auxpdm-counting} and
\Cref{thm:local-read-k} provide exact
$\Logspace$-uniform $\#\SAC^1$ circuits that compute the acceptance
counts of probabilistic auxiliary pushdown machines and of locally
read-$k$ machines, respectively.
Applying \Cref{thm:er-high-precision} to these circuits
yields approximations to their acceptance probabilities.

The two models can be handled together as follows.
In both cases there is a logspace-computable polynomial bound
$m(n)$ on the number of random bits such that
\[
 p_M(x)=\frac{F_M(x)}{2^{m(n)}},
\]
where $F_M(x)$ is the number of accepting random strings of length
$m(n)$. Thus an approximation to $F_M(x)$ with relative error
$2^{-t}$ gives, after division by $2^{m(n)}$, an approximation to
$p_M(x)$ with the same relative error.

\begin{corollary}
\label{cor:er-probabilities}
Let $M$ be either a polynomial-time probabilistic auxiliary
pushdown machine with $O(\log n)$ work space, or a polynomial-time locally read-$k$ machine
for any fixed $k$.
Let $p_M(x)$  be its acceptance probability.
For every fixed $a>0$ and every logspace-computable
integer-valued function $t=t(n)$ with $1\le t(n)\le n^a$, there is a polynomial-size, $\Logspace$-uniform family of Boolean $\AC$ circuits with
depth $ O\left(\log n + \frac{\log n}{\log\log n}\log\left(2+\frac{t}{\log n}\right) \right)$, that outputs $\widetilde p_M(x)$ satisfying
\[
 (1-2^{-t})p_M(x)\le\widetilde p_M(x)\le p_M(x),
 \qquad
 |\widetilde p_M(x)-p_M(x)|\le2^{-t}.
\]
The output is a $(t+3,O(\log n))$-number as in
\Cref{def:floating-point}: a floating-point number with
at most $t+3$ significand bits and an $O(\log n)$-bit signed
exponent.
In particular, for every fixed polylogarithmic $t(n)$, precision
$2^{-t(n)}$ is attainable in $\Logspace$-uniform $\AC^1$ without
assuming a bounded-error promise.
\end{corollary}

\begin{proof}
For auxiliary pushdown machines,
\Cref{lem:auxpdm-counting} gives
$F_M\in\Logspace\text{-}\#\SAC^1$ and
$p_M(x)=2^{-T(n)}F_M(x)$.
For locally read-$k$ machines,
\Cref{thm:local-read-k} gives
$F_M\in\Logspace\text{-}\#\SAC^1$,
and $p_M(x)=2^{-R(n)}F_M(x)$ by definition.
In either case, let $m(n)$ denote the corresponding
logspace-computable, polynomially bounded random-word length,
namely $T(n)$ or $R(n)$.

Apply \Cref{thm:er-high-precision} to $F_M$ with
$K(n)=\Theta(\log n)$, and let $\widetilde F_M(x)$ be the
resulting approximation.
Set
\[
 \widetilde p_M(x)=2^{-m(n)}\widetilde F_M(x).
\]
For nonzero outputs, multiplying by \(2^{-m(n)}\) subtracts
\(m(n)\) from the exponent; zero remains zero.
This rescaling is exact, preserves the significand, and leaves
the exponent length \(O(\log n)\).
It is computable by $\Logspace$-uniform circuits of depth
$O(\log n)$, which is within the $O(d(n,t))$ bound.
Thus rescaling preserves the relative-error guarantee.
Finally,
\[
 0\le p_M(x)-\widetilde p_M(x)
 \le 2^{-t}p_M(x)\le2^{-t}.
\]
\end{proof}

Taking $t(n)$ to be polylogarithmic gives the
acceptance-probability guarantees in
\Cref{thm:intro-auxpdm-approximation} and \Cref{thm:intro-local-read-approximation}.
Together with \Cref{thm:local-read-k}, this completes the proof
of \Cref{thm:intro-local-read-approximation}.

The acceptance-probability approximation also yields
decision simulations for smaller acceptance gaps.

\begin{corollary}
\label{cor:er-small-gap}
Suppose a machine $M$ from either model above satisfies
$p_M(x)\ge1/2+\gamma(n)$ on yes-instances and
$p_M(x)\le1/2-\gamma(n)$ on no-instances, where $\gamma(n)>0$.
If a logspace-computable integer-valued function $t$ with
$1\le t(n)\le n^a$ for some fixed $a>0$ satisfies
$2^{-t(n)}\le\gamma(n)/2$, then the language of $M$ is decided by
polynomial-size, $\Logspace$-uniform Boolean $\AC$ circuits of
depth $O(d(n,t))$.
\end{corollary}

\begin{proof}
Apply \Cref{cor:er-probabilities} and compare the output with
$1/2$.
On yes-instances it is at least $1/2+\gamma(n)/2$, while on
no-instances it is at most $1/2-\gamma(n)$.
The comparison can be performed in constant Boolean depth.
\end{proof}

\section{Proofs of Numerical Lemmas}
\label{sec:numerical-primitives}

We prove the four numerical lemmas stated in \Cref{subsec:floating-point-arithmetic}. Throughout this section, $m,p,b,q$ are positive, logspace-computable, polynomially bounded integer-valued functions of $n$.

The proof of approximate summation uses the following gap-majority problem and a result of Cheng and Wang \cite{ChengWang2024}.

\begin{definition}[Gap Majority]
Let $0\le a<b\le B$. The promise problem $\GapMaj[a,b]$ takes as input a Boolean string $y\in\{0,1\}^B$. Writing $|y|$ for its Hamming weight, the circuit must output $0$ if $|y|\le a$ and $1$ if $|y|\ge b$.
\end{definition}

\begin{theorem}[{\cite[Theorem 14]{ChengWang2024}}]
\label{thm:cheng-wang-gapmaj}
Let $B,a,b$ be integers with $B\ge4$ and $0\le a<b\le B$. There is a polynomial-size, $\Logspace$-uniform Boolean $\AC$ circuit for $\GapMaj[a,b]$ on $B$ input bits of depth
\[
O\left(1+\frac{\log\left(b/(b-a)\right)}{\log\log B}\right).
\]
Given $(1^B,a,b)$, a deterministic machine outputs a description of the circuit using $O(\log B)$ work space.
\end{theorem}

\elementaryArithmetic*
\begin{proof}
Directed rounding is trivially computable in constant depth. We also use the standard fact that addition, subtraction, and comparison of polynomial-length signed integers admit polynomial-size, $\Logspace$-uniform $\AC^0$ circuits.

Suppose $b=O(\log n)$. Write the two nonzero inputs as $A=M_A2^{e_A-\ell_A}$ and $B=M_B2^{e_B-\ell_B}$, and let $d=\min\{e_A-\ell_A,e_B-\ell_B\}$. Since the exponent values have magnitude $2^{O(b)}=n^{O(1)}$, both inputs can be aligned to the common scale $2^d$ by appending polynomially many zero bits to their significands. Exact addition and the nonnegative truncated difference then reduce to addition, subtraction, and comparison of two polynomial-length integers. Removing trailing zeros and adjusting the exponent, if needed, are constant-depth bit-selection operations.

Finally, if $p=O(\log n)$, the exact product of the two significands has $O(\log n)$ output bits. Its truth table has polynomial size, so each output bit can be computed by a constant-depth DNF whose wiring is generated in logspace. The exponent of the product is obtained by adding the two input exponents and adjusting for the significand lengths, all in $\Logspace$-uniform $\AC^0$. Zero inputs are handled by their flags. The output bounds follow from the polynomial input lengths and, for addition and subtraction, the polynomial bound on the exponent values.
\end{proof}

\roundedAddSub*
\begin{proof}
We prove rounded addition; rounded truncated subtraction follows by the same argument, with the boundary cases described below. Zero inputs are immediate. Write $A=M_A2^{e_A-\ell_A}$ and $B=M_B2^{e_B-\ell_B}$, and assume $e_A\ge e_B$. Put $d=e_A-e_B$ and $T=p+q+4$.

If $d\le T$, the two significands can be aligned exactly to a common power of two using words of $O(p+q)$ bits. Indeed, the difference between the lowest occupied bit positions is at most $d+p$, so a polynomial-length word contains both shifted significands. We compute their exact sum using constant-depth binary addition and then perform directed rounding to $q$ bits. All shifts, the selection of significant bits, and exponent adjustments have polynomial-size, $\Logspace$-uniform constant-depth implementations.

Suppose instead that $d>T$. Since $A\in[2^{e_A-1},2^{e_A})$ and $B<2^{e_B}$, we have
\[
0<B<2^{e_A-p-q-4}.
\]
Both $A$ and every $q$-bit representable number in the neighborhood of $A$ are integer multiples of powers of two with exponent at least $e_A-\max\{p,q\}-1$. Consequently, the distance from $A$ to any distinct neighboring $q$-bit representable number is at least $2^{e_A-\max\{p,q\}-1}$, which exceeds $B$. Therefore
\[
\operatorname{RoundDown}_q(A+B)=\operatorname{RoundDown}_q(A),
\]
while $\operatorname{RoundUp}_q(A+B)$ equals $\operatorname{RoundUp}_q(A)$ unless $A$ is already representable with at most $q$ significant bits, in which case it is the immediate $q$-bit successor of $A$. Representability and the immediate successor can be determined directly from the significand in constant depth. This computes both rounded sums without constructing the potentially very long exact sum.

For subtraction, if $A\le B$ both outputs are zero. Otherwise, when $|e_A-e_B|\le T$, align the two inputs as above, subtract the resulting polynomial-length integers, and round. If $e_A-e_B>T$, the smaller number cannot cross a rounding boundary except when $A$ itself is exactly representable with at most $q$ significant bits. In that case $\operatorname{RoundDown}_q(A-B)$ is the immediate $q$-bit predecessor of $A$ and $\operatorname{RoundUp}_q(A-B)=A$; otherwise both rounded differences agree with the corresponding rounding of $A$. These cases are decided in constant depth. In every case the output exponent differs from one of the input exponents by at most $O(p+q)$, so it has $b+O(\log n)$ bits. All constructions have polynomial size and logspace-uniform wiring.
\end{proof}

\exactArithmetic*
\begin{proof}
We first recall the Boolean implementation of exact arithmetic on polynomial-length integers. Let $W=n^{O(1)}$ bound the operand lengths and $m=n^{O(1)}$ the number of summands. A carry-save adder reduces three binary rows to two in constant bounded-fan-in depth. Repeating this reduction for $O(\log(m+1))$ levels leaves two rows, whose sum is computed by a parallel-prefix adder in depth $O(\log(W+\log(m+1)))$. Thus exact iterated addition has polynomial size, $\Logspace$-uniform wiring, and bounded-fan-in depth $O(\log n)$. Multiplication reduces to the exact sum of the partial-product rows and has the same bounds.

To obtain the improved Boolean depth, replace each constant-fan-in gate by a constant-size fan-in-two circuit, layer the result, and group $h=\lfloor\log\log n\rfloor$ consecutive layers into each block. The output of a block depends on at most $2^h\le\log n$ boundary-input occurrences. Its Boolean function therefore has an exact DNF of at most $2^{2^h}\le n$ terms and constant depth. Replacing each block by this DNF preserves polynomial size and gives depth $O(\log n/\log\log n)$. The replacement is logspace-uniform: a generator enumerates the assignments to the boundary inputs and evaluates each unfolded fan-in cone in $O(\log n)$ space, recovering internal gate identifiers from their paths and the original logspace-uniform wiring. The finitely many small values of $n$ are handled separately.

For multiplication of $(p,b)$-numbers, compute the exact product of the two polynomial-length significands and add their signed exponents. The product significand is odd, has at most $2p$ bits, and its normalized exponent has at most $b+2$ bits. For summation with $b=O(\log n)$, align all nonzero inputs to the least occupied bit position. Their exponent values differ by less than $2^b$, so the aligned integers have at most $p+2^b+O(1)$ bits. The exact sum consequently has at most $p'=p+2^b+\lceil\log m\rceil+2$ bits. After removal of trailing zeros, its exponent has at most $b+O(\log\log n)$ bits. Alignment, conversion to normalized representation, and the exponent calculations have constant Boolean depth; the iterated integer sum has depth $O(\log n/\log\log n)$ as proved above. All word lengths and connections are bounded and generated in logspace.
\end{proof}

\positiveSummation*
\begin{proof}
Put $Q=q+6$ and $\eta=2^{-Q}$. Since all summands are nonnegative, $S=0$ exactly when every input is zero, which is testable in constant depth. In this case output $L=U=0$. Assume $S>0$, and choose the integer $E$ such that
\[
2^{E-1}\le\max_i A_i<2^E.
\]
The exponent $E$ is selected from the nonzero input exponents by parallel comparisons. Set
\[
g=Q+\lceil\log m\rceil+4,\qquad \alpha=2^{E-g},\qquad c_i=\left\lfloor\frac{A_i}{\alpha}\right\rfloor,\qquad C=\sum_{i=1}^m c_i.
\]
Each $c_i$ is a nonnegative integer of at most $g$ bits. It is obtained by comparing the exponents and selecting a window of the input significand, without expanding an exponent-sized binary word; this takes constant Boolean depth even when $p,b$ are polynomially large. Since $0\le A_i-\alpha c_i<\alpha$ and $S\ge2^{E-1}$, we have
\begin{equation}
\alpha C\le S<\alpha(C+m),\qquad \frac{m\alpha}{S}\le 2m2^{-g}\le\frac{\eta}{8}.
\label{eq:numerical-sum-truncation}
\end{equation}

We distinguish two precision ranges. If $q\le\lceil\log n\rceil$, then $g=O(\log n)$ and $C$ can be encoded as the Hamming weight of a polynomial-length Boolean string. Specifically, for every $i$ and $1\le j\le2^g$, use the bit $[j\le c_i]$. The resulting string has length $B_0=m2^g=n^{O(1)}$ and weight exactly $C$. Pad it with zeros to a logspace-computable length $B\ge\max\{4B_0,n\}$. For every $1\le\lambda\le B_0$, run in parallel the gap-majority circuit of \Cref{thm:cheng-wang-gapmaj} with thresholds
\[
a_\lambda=\lambda,\qquad b_\lambda=\lceil(1+\eta)\lambda\rceil.
\]
Its depth is $O(1+Q/\log\log B)$, because $b_\lambda/(b_\lambda-a_\lambda)=O(\eta^{-1})$. Let $\lambda_0$ be the smallest index whose test outputs zero. The test at $\lambda=C$ outputs zero, whereas a zero output at $\lambda_0$ is impossible when $C\ge b_{\lambda_0}$. Thus
\[
\lambda_0\le C<b_{\lambda_0}\le(1+\eta)\lambda_0+1,
\qquad\text{and hence}\qquad
\lambda_0\le C<(1+\eta)\lambda_0.
\]
Take
\[
L_0=\alpha\lambda_0,\qquad U_0=\alpha\bigl((1+\eta)\lambda_0+m\bigr).
\]
By \eqref{eq:numerical-sum-truncation} and $\alpha\lambda_0\le S$, these are lower and upper bounds for $S$ with relative errors at most $\eta+\eta/8$. Compute $L=\operatorname{RoundDown}_Q(L_0)$ and $U=\operatorname{RoundUp}_Q(U_0)$. Each rounding adds relative error at most $2^{1-Q}=2\eta$, so the total error is less than $5\eta<2^{-q}$. The Boolean operations outside the parallel gap-majority calls have constant depth, and all test lengths and thresholds are logspace-computable. The total depth in this case is $O(1+q/\log\log n)$.

If $q>\lceil\log n\rceil$, then $g$ is polynomially bounded. Compute $C=\sum_i c_i$ exactly using the polynomial-length integer-summation construction from the proof of \Cref{lem:er-arithmetic}, in depth $O(\log n/\log\log n)$. Put $L_0=\alpha C$ and $U_0=\alpha(C+m)$, then round downward and upward to $Q$ significant bits. Equation~\eqref{eq:numerical-sum-truncation} and the rounding bound again give relative errors smaller than $5\eta<2^{-q}$. The additional operations have constant depth, so the total depth is $O(\log n/\log\log n)$.

In both cases the outputs have at most $Q=q+6$ significant bits. Since $2^{E-1}\le S<m2^E$ and the approximations are within a factor of two of $S$, their exponents differ from $E$ by at most $O(\log m+1)$. Thus $b+O(\log\log n)$ exponent bits suffice. Combining the two precision ranges gives depth $O(1+\min\{q,\log n\}/\log\log n)$, polynomial size, and $\Logspace$-uniformity.
\end{proof}

\section*{Acknowledgements}
We thank William M. Hoza for helpful conversations on this work.

\paragraph{Statement on the Use of AI}

 We formulated the research topics and questions of this work without using AI and had the general solution framework in mind before using AI. We used ChatGPT (GPT-6 Astra) to assist with proof development and proof checking, making substantial use of the tool for both tasks. The tool materially affected the proofs throughout all technical sections of the paper. The manuscript was written by the authors with assistance from the same tool. The authors take full responsibility for the correctness of the work.


\appendix

\section{Approximation with the Circuit as Input}
\label{subsec:input-circuit-approximation}

This section considers a side question whether our construction can be ``universal''. Here ``universal'' is in the sense that we can construct $\Logspace\text{-}\AC^1$ circuits that takes arithmetic circuits and their inputs as the input, and compute the evaluation. 

\begin{corollary}[Approximating Positive Arithmetic Circuits Given as Input]
\label{cor:input-circuit-approximation}
$K(N),t(N)\ge1$ are logspace-computable, polynomially bounded integer-valued functions. There is a polynomial-size, $\Logspace$-uniform Boolean $\AC$ circuit family of depth $O(K(N)+t(N))$ that takes as input an explicit description of a positive semi-unbounded arithmetic circuit $G$ and a Boolean assignment $x$. Let $N\ge2$ be the total input length, and suppose that $G$ has depth at most $K(N)$. The circuit family outputs a $(t(N)+2, O(K(N)+\log N))$-number $\widetilde F(G,x)$ in floating-point representation, satisfying
\[
    (1-2^{-t(N)}) G(x)
    \le \widetilde F(G,x)
    \le G(x).
\]
In particular, if $K(N),t(N)=O(\log N)$, this approximation is computable by a $\Logspace$-uniform $\AC^1$ circuit family.
\end{corollary}

\begin{proof}
We first decode $(G, x)$ using a $\Logspace$-uniform Boolean $\AC^0$ circuit family, then simulate $G$ by a positive semi-unbounded arithmetic circuit whose inputs are the decoded bits. Fix a padded encoding in which gate types, input labels, incoming-wire occurrences, and the designated output are stored in fixed-length fields. Equality between an $O(\log N)$-bit field and a fixed label is an AND of Boolean literals and can therefore be computed in constant Boolean depth.

Let $a_v,b_v$ indicate whether gate $v$ is an addition or multiplication gate, respectively, and let $e_{v,h,j}$ indicate whether its $h$th input wire comes from gate $j$. Unused input slots have zero indicators, and parallel wires occupy distinct slots. Let $u_v$ be the value of the literal or constant at an input gate under the assignment $x$, and let $u_v=0$ at an internal gate. Finally, let $o_v$ indicate whether $v$ is the designated output gate. All these bits can be computed in constant Boolean depth and polynomial size from the encoding and $x$. Write $y(G,x)$ for the resulting bit vector.

We construct a positive semi-unbounded arithmetic circuit $U_N$ that takes these decoded bits as inputs. Set $D=K(N)$. Starting with $z_v^{(0)}=u_v$, for each $0\le r<D$ define
\[
\begin{aligned}
    R_{v,h}^{(r)}
        &=\sum_j e_{v,h,j}z_j^{(r)},\\
    z_v^{(r+1)}
        &=u_v+a_v\sum_h R_{v,h}^{(r)}
          +b_vR_{v,1}^{(r)}R_{v,2}^{(r)}.
\end{aligned}
\]
Each round uses a constant number of layers of addition gates and binary multiplication gates. On a valid encoding, $R_{v,h}^{(r)}$ selects the current value of the gate supplying the $h$th input wire of $v$. The update therefore applies the operation specified by $v$, while preserving the values of input gates.

Induction on gate depth shows that $z_v^{(r)}$ equals the value computed at gate $v$ whenever $r$ is at least its depth. Since $G$ has depth at most $D$, we obtain
\[
    U_N(y(G,x))
    =\sum_v o_vz_v^{(D)}
    =G(x).
\]

The circuit $U_N$ has polynomial size and depth $O(K(N))$. Its gates and wires are indexed by rounds, gate indices, and input slots, all of which use $O(\log N)$ bits because $K(N)$ is polynomially bounded. A logspace transducer can enumerate these indices and generate the corresponding connections, so the family $(U_N)_{N\ge2}$ is $\Logspace$-uniform.

Pad the input vector of $U_N$ to length $M=N^d$ for a fixed integer $d$. Reindexing this family by $M$ allows us to apply \Cref{thm:approx-sharp-sac} with accuracy parameter $t(N)$; at input lengths that are not perfect $d$th powers, use a constant-zero circuit. Since $N$ is recoverable from $M$ in logspace and $\log M=O(\log N)$, the theorem gives a polynomial-size, $\Logspace$-uniform Boolean circuit family of depth $O(K(N)+t(N))$, with the claimed approximation and representation bounds. Composing this family with the constant-depth decoding circuit proves the corollary.
\end{proof}

The uniformity requirement applies to the decoding circuit family, the universal arithmetic circuit family, and the resulting Boolean approximation circuit family. The input circuit $G$ need not belong to a $\Logspace$-uniform family.

\bibliographystyle{alpha}
\bibliography{refer}

\end{document}